# Reassessing carotenoid photophysics: shedding light on dark states

Roxanne Bercy, Viola D'mello, Andrew Gall, Cristian Ilioaia, Andrew A. Pascal, Juan Jose Romero*, Bruno Robert*, Manuel J. Llansola-Portoles*

Université Paris-Saclay, CEA, CNRS, Institute for Integrative Biology of the Cell (I2BC), 91198, Gif-sur-Yvette, France.

*KEYWORDS. Carotenoids, dark excited states, femtosecond stimulated resonance Raman spectroscopy.*

**ABSTRACT:** Carotenoid molecules are critical in photosynthesis, performing functions at the heart of both light-harvesting and photoprotection. As both these processes involve excitation energy transfer, fully understanding them requires a precise description of the electronic states involved. The excited state manifold of carotenoids is not yet fully characterised, and includes several dark electronic states that remain elusive. Using femtosecond stimulated resonance Raman spectroscopy, where the vibrational contributions of each excited state can be observed selectively as a function of the Raman excitation, we resolve vibrational signatures consistent with three additional dark-state contributions and propose assignments for them. These results address long-standing controversies in carotenoid research and provide a spectroscopic framework relevant to the multiple roles of these molecules.

## Introduction

Carotenoids are ubiquitous pigments that perform diverse natural functions. These range from colour-based signalling, and protection against reactive oxygen species, to essential roles in photosynthesis - harvesting solar energy and protecting the photosynthetic apparatus from photodamage. However, our understanding of carotenoid photophysics, and hence of light-driven processes in carotenoid-containing systems, remains largely incomplete - despite intensive investigation over many decades. The key electronic states of carotenoids have been described by a simplified three-state model - comprising the ground state $S_0$ ($1^1A_g^-$), the "dark" $S_1$ state ($2^1A_g^-$, silent in one-photon absorption), and the optically bright $S_2$ state ($1^1B_u^+$) responsible for the intense 450–550 nm absorption band. Nonetheless, carotenoid excited-state dynamics extends far beyond this simple model, and additional transient features have been observed, leading to schemes that include several additional "dark" states ($S_x$, vibrationally hot $S_1$, S*, and an internal charge-transfer state, ICT), without any clear consensus on their assignment [1-14]. Ultrafast techniques such as transient absorption (TA) [3, 15] and femtosecond time-resolved stimulated Raman spectroscopy (FSRS) [4, 11, 14] are key approaches for probing carotenoid dark states, but the spectral congestion inherent in the former TA, and the absence of selectivity in the latter FSRS, have so far impeded unambiguous assignments.

To overcome the limited state selectivity of TA and conventional FSRS, we have used femtosecond stimulated resonant Raman spectroscopy (FSRRS), which combines ultrafast pump–probe timing with the mode specificity of Raman scattering under resonance enhancement [16-18]. In a standard FSRS sequence, an actinic pump (≈100 fs) prepares the excited-state population. After a controlled delay, a spectrally narrow Raman pump (RP; ≈3 ps, ≈5 cm$^{-1}$) and a broadband probe (≈100 fs) are temporally and spatially overlapped in the sample. Their interaction generates stimulated Raman gain and loss features in the probe spectrum at well-defined vibrational shifts relative to the RP frequency, providing a vibrational fingerprint of the transient species present at that delay. In FSRS implementations, the RP is fixed in the near-IR and is therefore off-resonant with the relevant electronic transitions, so signals from multiple coexisting states can overlap. In FSRRS, the RP is fully tunable into the visible and can be chosen in resonance with an electronic transition of a targeted excited species. Under resonance, the Raman cross-section of that species is strongly enhanced, so its vibrational bands dominate the stimulated Raman response while off-resonant contributions are comparatively suppressed, enabling state-selective vibrational tracking. Although resonance enhancement is well established, systematic datasets that exploit tunable resonance conditions to disentangle a dense manifold of overlapping dark states, as encountered in carotenoids, have not yet been recorded. The resulting signal-to-noise and spectral selectivity allow the kinetics of some individual vibrational bands to be followed directly, enabling a more robust separation and assignment of overlapping transient contributions.

We have applied this method to a series of linear carotenoids with increasing conjugation length - neurosporene, lycopene, and spirilloxanthin (9, 11, and 13 effective conjugated C=C bonds ($N_{eff}$), respectively) [19-20]. To aid the assignment of the spectral features observed in linear carotenoids, we performed complementary measurements on two cyclic carotenoids: β-carotene, which contains two conjugated ionone rings, and fucoxanthin, which is known to exhibit a well-characterised intramolecular charge transfer (ICT) state. Within our time resolution, these data allowed us to track the formation and decay of all observed components, and indicate that the carotenoid excited state manifold generally comprises at least three optically dark states. By comparing their timescales and spectral signatures across the different systems, we propose assignments and symmetry labels for each state.

## Results and Discussion

**Electronic Transition Fingerprints**.

Figure 1 shows the *all-trans* structures of neurosporene (9 conjugated C=C bonds), lycopene (11 bonds), and spirilloxanthin (13 bonds), with their π-conjugated backbones highlighted, alongside their optical signatures at 298 K. The steady-state absorption spectra (black lines) exhibit the characteristic three-band vibronic progression, with 0–0 transitions at 475 nm (neurosporene), 512 nm (lycopene), and 532 nm (spirilloxanthin), in agreement with extensive literature reports [20-21]. Gaussian fits to the 0–0 band yield full widths at half maximum (FWHM) of 788, 965, and 1018 cm$^{-1}$ for neurosporene, lycopene, and spirilloxanthin, respectively. The systematic broadening is consistent with increased conformational disorder or flexibility. To assess possible cis isomer content responsible for this broadening, we recorded resonance Raman spectra under 0–0 excitation and observed no modes in the 1120–1150 cm$^{-1}$ region [22-24], supporting a predominantly all-trans configuration (Fig. S1 in Supporting Information).

We first used TA to establish the excited state spectral signatures and, in turn, to define the Raman pump wavelength range required for tuning the Raman pump in the FSRRS measurements (*vide infra*). Figure 1 presents time-gated TA spectra at selected delays, with main ESA maxima at 500 nm (neurosporene), 572 nm (lycopene), and 604 nm (spirilloxanthin) (full datasets in figure S2, Supporting Information). The main ESA band also broadens with increasing conjugation length. Because multiple excited state contributions evolve in parallel, a quantitative linewidth is not straightforward to extract; nevertheless, the qualitative trend is consistent with increased conformational flexibility. Kinetic traces extracted at the ESA maximum and at the blue and red shoulders are shown in the side panels. For all three carotenoids, the blue feature (commonly labelled S*) and the rapidly decaying red-shifted shoulder appear within the instrument response (circa 150 fs). In contrast, the main ESA band, assigned primarily to $S_1$, rises on a timescale comparable to the decay of the red shoulder. This behaviour is consistent with previous reports in which the dominant ESA is attributed to the $S_1$–$S_n$ transition, with $S_1$ lifetimes of 22, 4.2, and 1.4 ps for neurosporene, lycopene, and spirilloxanthin, respectively. In each case, the $S_1$-associated spectrum is accompanied by a short-lived red shoulder (hundreds of femtoseconds) and a blue-shifted component persisting into the picosecond regime. The physical origin of both shoulders remains debated: the red component has been attributed either to an additional excited state (often denoted $S_x$) or to vibrationally hot $S_1$, whereas the blue component (S*) has been variously interpreted as a distinct electronic state or as a vibrationally hot ground-state response. As a result, neither its symmetry nor its spectroscopic character is currently established.

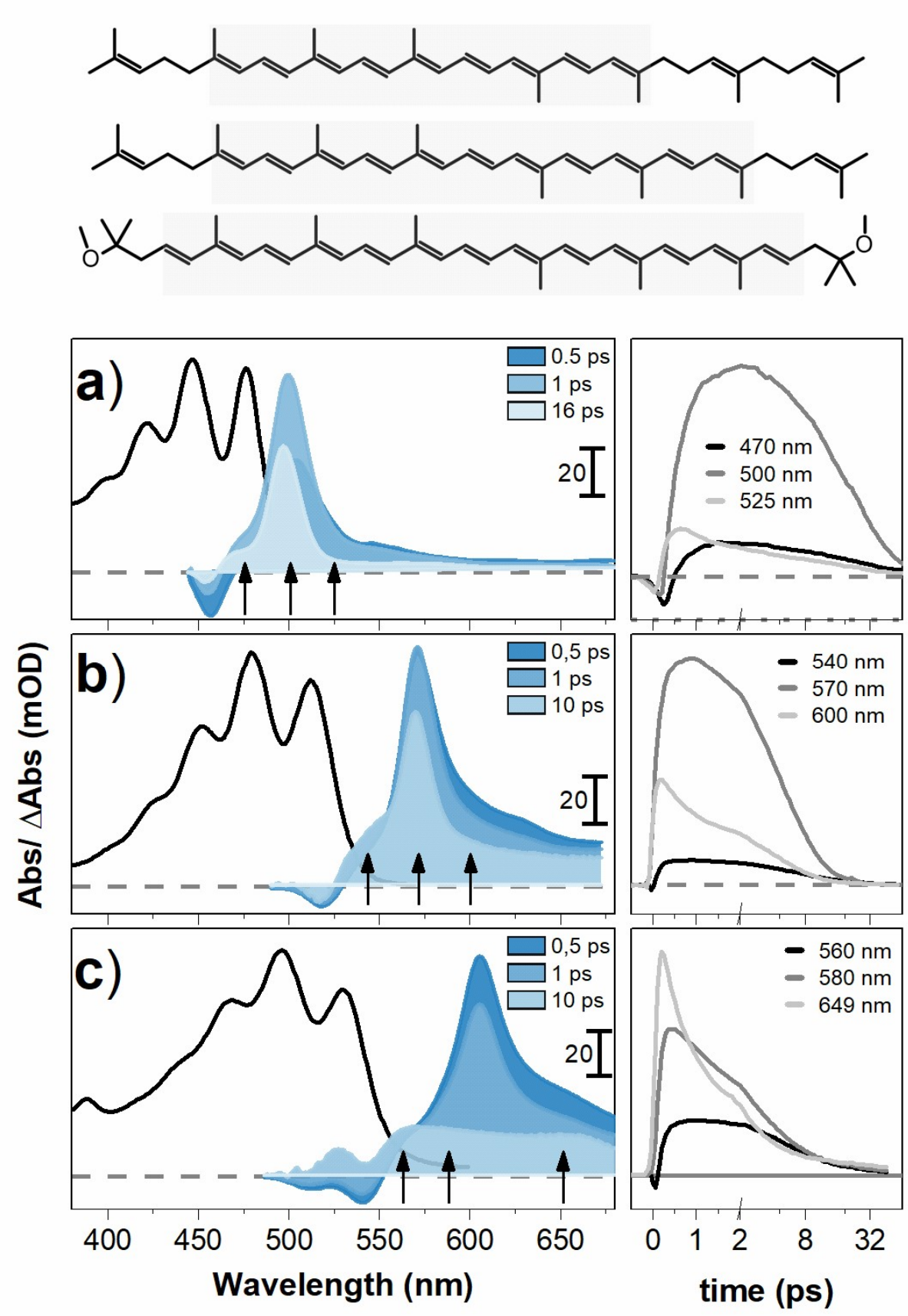

**Figure 1**. Molecular structures of all trans neurosporene (top), lycopene (middle), and spirilloxanthin (bottom). Panels show **(a)** neurosporene (n hexane), **(b)** lycopene (THF), and **(c)** spirilloxanthin (THF). Each panel includes the steady state absorption spectrum (black) and time gated transient absorption (TA) spectra (colour shaded regions); the TA amplitude scale (mOD) is given by the adjacent bar. Kinetic traces at probe wavelengths marked by black arrows are shown on the left. All measurements were recorded at room temperature.

**Resonance Raman fingerprints of the excited states**. Guided by the spectral features observed in TA, we then chose a range of RP wavelengths for FSRRS, matching this stimulated absorption region for each carotenoid, generating sets of time/Raman-shift colourmaps. Figure 2 illustrates this, by displaying a selection of such maps at different wavelengths of the Raman pump for the three main carotenoids under study (β-carotene and fucoxanthin maps are shown in Supplementary Information). In these spectra, C=C stretching modes contribute between 1400 and 1840 cm$^{-1}$ ($\nu_1$ region), stretching vibrations of C-C single bonds coupled with C-H in-plane bending modes form an envelope spanning 1100–1310 cm$^{-1}$ ($\nu_2$ region), and in-plane rocking vibrations of the methyl groups contribute around 1000-1100 cm$^{-1}$ ($\nu_3$ region).

The *resonance* conditions used here ensure intense Raman signals, such that it is possible to extract time-gated Raman spectra from these maps at selected delays, together with kinetic traces for each frequency. For all carotenoids studied, four components are observed in the $\nu_1$ region, denoted

$\nu_{1a}$ to $\nu_{1d}$ and contributing at 1770-1800, 1730-1760, 1530-1580 and 1475-1505 cm$^{-1}$, respectively. Note that the earliest times after the actinic pulse should be treated with some care, the time resolution being limited by the instrument response function (< 150 fs). In addition, bleaching of the intense ground-state Raman bands at time zero (blue features in Fig. 2), itself modulated by the coherent artefact (due to coincidence of three pulses at t=0; ±75 fs timescale), distorts the less intense excited-state features in their vicinity. Distinct vibrational modes are preferentially enhanced depending on the RP wavelength, so that the contributions of individual excited states can be extracted from such maps. Besides some obvious features, such as the strong contribution around 1790 cm$^{-1}$ widely described in the literature for the S$_1$ state [4], the neurosporene map at 530 nm exhibits a strong but short-lived mode at 1758 cm$^{-1}$, accompanied by a broad smear around 1200-1300 cm$^{-1}$ and a collection of low intensity features at 1030, 1130, 1475 and 1570 cm$^{-1}$ (visible from ~0.25 ps). Lycopene and spirilloxanthin exhibit similar behaviours, but the modes *circa* 1030 and 1490 cm$^{-1}$ decay more slowly than the other modes. Note that in long carotenoids the higher energy Raman pumps (to the blue) preferentially enhance the 1030, 1130 and 1470-1505 cm$^{-1}$ modes, whereas in neurosporene these modes are enhanced under redder excitations. Another common feature across the three carotenoids is the preferential enhancement of the $\nu_1$b mode. However, it is difficult to determine whether the apparent evolution reflects a gradual shift in intensity from $\nu_{1b}$ to $\nu_{1a}$ or, instead, a decay of one mode accompanied by the rise of the other.

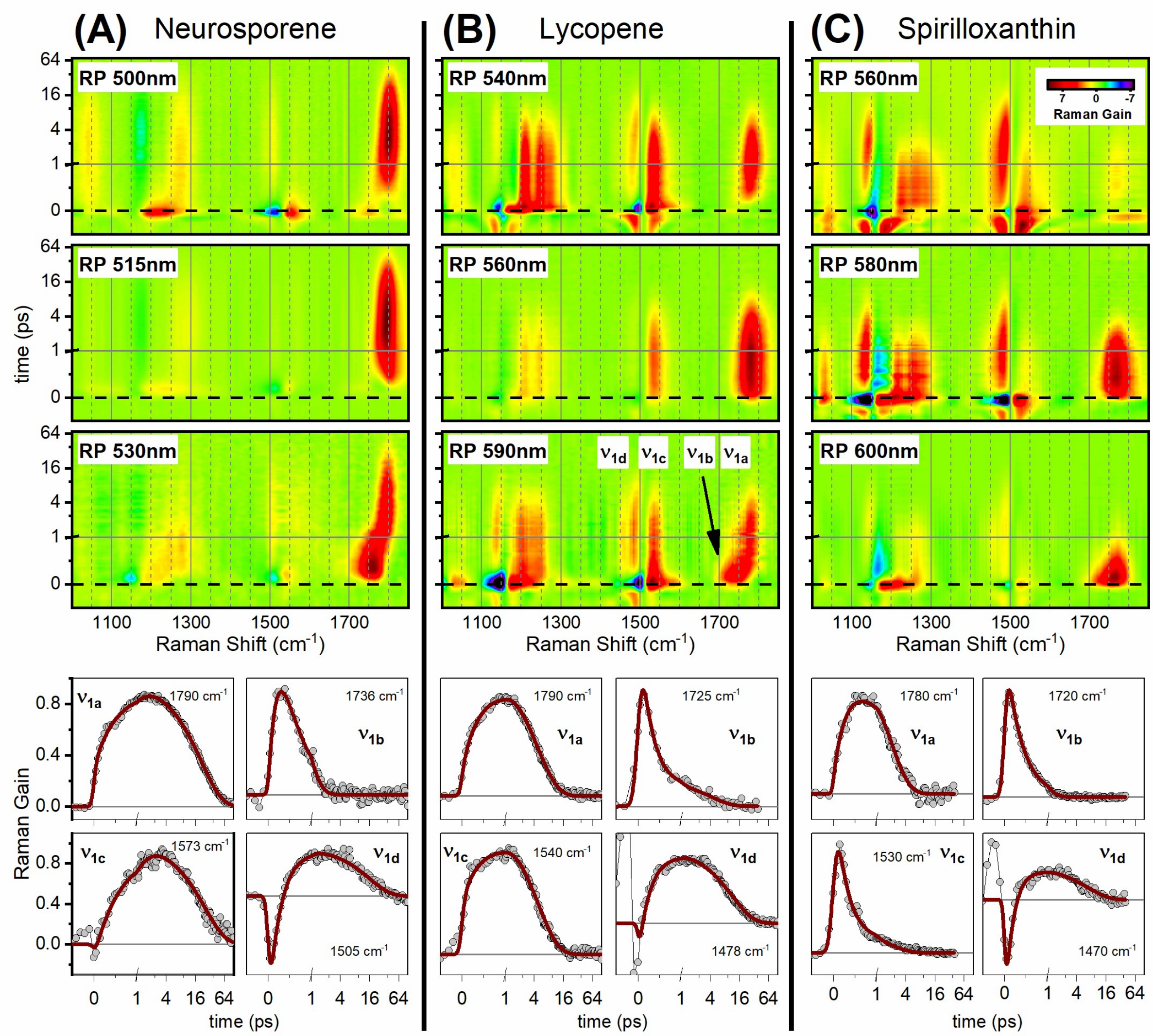

**Figure 2**. Femtosecond Stimulated Resonance Raman of carotenoids at room temperature - time-spectral maps and kinetics. Colour-maps of Raman intensity as a function of Raman shift (cm$^{-1}$) and pump–probe delay (linear scale to 1 ps, logarithmic thereafter) for: (A) neurosporene in n-hexane; actinic pump 475 nm, Raman pump at 500, 515 & 530 nm; (B) lycopene in THF; actinic pump 510 nm, RP 540, 560 & 590 nm; (C) spirilloxanthin in THF; actinic pump 540 nm, RP 560, 580 & 600 nm. For simplicity, the $\nu_{1a}$-$\nu_{1d}$ regions are only indicated in panel (B) RP 590 nm, but they are common to all carotenoids at all excitations. Lower panels show the selected pump–probe delays and representative kinetic traces (with exponential fits) for each of the $\nu_1$(a-d) modes. The gated spectra of the time spectral maps are shown in figure S3. The full set of time–spectral maps for all RP wavelengths is provided in figure S4-6 in Supplementary Information.

Performing individual fittings of the time traces of each of these Raman modes allows their specific formation and decay rates to be determined (Table 1; kinetic fittings in Fig. 2). $\nu_{1c}$ at 1550 $cm^{-1}$ may already be present at time zero (with the *proviso* mentioned above, regarding the earliest timescales), while modes $\nu_{1b}$ and $\nu_{1d}$ rise within the instrument response time (*circa* 150 fs), but decay over very different timescales (< 1 ps and 5-20 ps, respectively) – indicating a branching of the kinetic pathway. Mode $\nu_{1a}$ then appears after a few hundred femtoseconds along with an increase in $\nu_{1c}$, and these two modes decay concomitantly. Note that the formation rate of $\nu_{1a}$ matches the decay rate of $\nu_{1b}$, while the origin of the increase in $\nu_{1c}$ is unclear due to its initial presence within the instrument response (< 150 fs). A potential concern in FSRRS is that the Raman pump itself may act as a secondary actinic source, perturbing the excited-state populations under study. To address this point, we compared transient absorption signals recorded with and without the Raman pump for each carotenoid at 0.5 and 5 ps (Fig. S15), and examined the dependence of the FSRRS spectra of lycopene on Raman pump fluence over the range 400 to 600 μJ $cm^{-2}$ (Figs. S16–S18). In all cases, the shape and time evolution of the signals are conserved, indicating that under our experimental conditions the Raman pump does not produce a measurable secondary actinic effect on the populations followed in this work.

**Table 1.** $\nu_1$-band centres and lifetimes. Each cell reports band frequency ($cm^{-1}$) / decay rate τ (ps, from single-exponential fit to corresponding kinetic trace). $N_{eff}$ indicated in parentheses for each carotenoid.

| Carotenoid ($N_{eff}$) | Solvent | $\nu_1$d ($cm^{-1}$) / τ (ps) | $\nu_1$c ($cm^{-1}$) / τ (ps) | $\nu_1$b ($cm^{-1}$) / τ (ps) | $\nu_1$a ($cm^{-1}$) / τ (ps) |
|---|---|---|---|---|---|
| Neurosporene (9) | *n*hexane | 1505 / 23.0 | 1575 / 22.0 | 1758 / 0.80 | 1797 / 21.9 |
| | THF | 1503 / 17.0 | 1570 / 23.1 | 1763 / 0.80 | 1795 / 23.0 |
| β-Carotene (9.6) | *n*hexane | — / — | — / — | — / — | — / — |
| | THF | 1496 / 10.8 | 1550 / 10.7 | 1765 / 0.50 | 1789 / 10.6 |
| Lycopene (11) | *n*hexane | 1488 / 8.5 | 1537 / 4.2 | 1740 / 0.40 | 1790 / 4.2 |
| | THF | 1486 / 9.0 | 1536 / 4.7 | 1744 / 0.56 | 1779 / 4.6 |
| Spirilloxanthin (13) | *n*hexane | 1475 / 6.9 | — / — | 1754 / — | 1770 / 1.6 |
| | THF | 1477 / 7.2 | 1533 / 1.8 | 1750 / 0.30 | 1772 / 1.6 |

*Notes:* "—" = not observed or not reliably determined under these conditions.

**Global modelling**. The FSRRS matrices were globally analysed to disentangle the vibrational modes associated with each of the excited states. As observed in figure 2, no single dataset contains all of the features defined perfectly well, as each RP enhances different features (as expected due to the selectivity inherent in resonance conditions). For this reason, we chose to show in figure 3A, the modelling of the dataset for lycopene with RP at 540 nm, which favours modes $\nu_{1a}$, $\nu_{1c}$, and $\nu_{1d}$. To show the consistency of our data, we show in the supporting information the modelling for the dataset of lycopene obtained at RP 590 nm, which enhances mode $\nu_{1b}$ (Fig. S10).

The parameters from TA and single-mode fits (modes $\nu_{1a}$-$\nu_{1d}$, Table 1) were used to seed the global model, and were then allowed to evolve freely. Simple sequential or parallel models were unable to resolve the spectral features satisfactorily, requiring rather the use of branched models – unsurprising, given the differential decays observed for $\nu_{1b}$ & $\nu_{1d}$ (see discussion above). Considering the spectral separation between the $\nu_1$ vibrational modes and the differences in formation and decay rates obtained by individual curve fitting, we propose a branched model (Fig. 3A) that fully describes the evolution of the carotenoid excited states, where a clear separation of the species-associated spectra (SAS) is achieved - shown for lycopene at RP=540 nm in Fig. 3B, and for the other linear carotenoids (RP=540 nm) in Supplementary information Figs S11-12. For comparison, the sequential and target models with 4 components are shown for the same lycopene dataset in the supporting information (figure S7-8). We resolve a consistent five-component model (ES1–ES5): **ES1** decays within the instrument response (< 150 fs), so that the vibrational signatures are convoluted with the IRF. Although fitting carries more uncertainty that the other components, it is still worth noting that it shows positive bands at $\nu_1 \approx 1550$ $cm^{-1}$ ($\nu_{1c}$) plus a broad $\nu_2$ feature near 1250 $cm^{-1}$, in line with recent reports by Polívka and Kloz[11]. **ES2** is formed from ES1 in less than 150 fs, within the instrument response, so we cannot determine the specific formation raise. ES2 lives sub-picosecond, with a lifetime that shortens from ~0.8 ps in neurosporene to ~0.3 ps in spirilloxanthin. It corresponds to $\nu_{1b}$ and exhibits positive bands near 1750 and 1250 $cm^{-1}$ (well-resolved in lycopene, probably due to favourable resonance). **ES2** feeds **ES3**, assigned to $\nu_{1a}$, accompanied by strong modes around 1250 $cm^{-1}$. **ES4,** assigned to $\nu_{1c}$, forms within < 150 fs and decays concomitantly with ES3 in a few picoseconds. **ES5** forms within < 150 fs and persists on the picosecond timescale - it corresponds to $\nu_{1d}$, lacks prominent higher-frequency features, and exhibits bands on the low-energy edge of $\nu_2$ (~1130 cm$^{-1}$). It is important to note that the formation of states ES2, ES4 & ES5 (a three-way branching from ES1 in the model) occurs within the experimental instrument response function (<150 fs), concealing a complex ultrafast dynamic process. Thus, while the temporal resolution is sufficient to characterise these states once formed and their following evolution, we cannot be sure of the underlying processes leading to their formation on the sub-150 fs timescale.

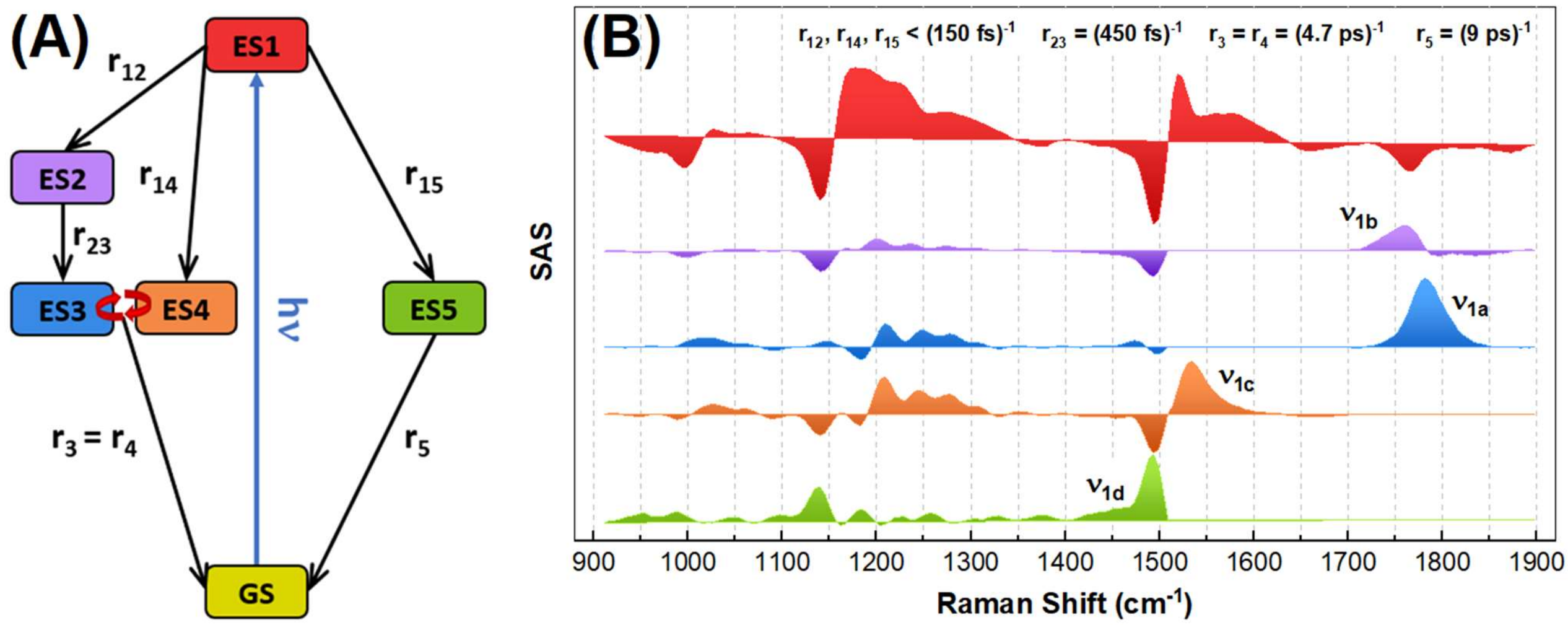

**Figure 3**. **Global kinetic analysis of carotenoid excited-state dynamics. (A)** The branched kinetic scheme is colour-coded by state (ES1-ES5), and the fitted values are reported as lifetimes ($\tau = 1/r$). **(B)** Species-associated spectra (SAS) are shown for lycopene in THF; actinic pump 510 nm, RP 540 nm. Full comparisons of experimental and modelled FSRRS matrices are provided in supporting information Fig S9**.**

The deconvolution achieved allows us to extract the precise frequencies of all four modes $\nu_{1(a\text{-}d)}$, which exhibit a clear linear relationship with the reciprocal of the effective conjugation length, $N_{eff}$ (Fig. **4A**), mirroring the behaviour of the ground-state $\nu_1$ mode [21]. Consistent with this structure-energy relationship, their decay rates also vary with conjugation length (Fig. **4B**). The decay rates for component ES3 (associated with $\nu_{1a}$) match literature values for $S_1$ in neurosporene, β-carotene, lycopene, and spirilloxanthin, respectively [4-5], and consequently follow the band-gap law (fig. 4A, blue fit) [25]. The decay rates for component ES5 and the positions of its $\nu_{1d}$ mode are consistent with literature values for S* in lycopene and spirilloxanthin [4,26]. The apparent resonance profiles of $\nu_{1d}$ are also consistent with S*, appearing somewhat more intense for the bluest RP in each case (S* absorption on the blue side of $S_1$). We therefore attribute component ES5 to S* in lycopene and spirilloxanthin - and by extension, it seems likely that $\nu_{1d}$ also reflects the presence of S* for neurosporene and β-carotene, even though this state has not been directly observed in these shorter carotenoids before. Following the trend for both components ES3 and ES5 (figure **4B**, blue and green exponential fits), the $S_1$ and S* rates intersect at $N_{eff} \approx 9.6$, so that for carotenoids shorter than β-carotene, S* absorption decays faster than $S_1$, and thus becomes masked by $S_1$ signatures in transient absorption. This explains very neatly why S* has never been observed before in shorter carotenoids using non-resonant techniques.

Whereas the $\nu_{1a}$ mode of component ES3 has been unequivocally linked to $S_1$ [4], the concomitant decay of mode $\nu_{1c}$ (component ES4) is more intriguing - there are no reports associating this frequency with the $S_1$ state. We compared the relative intensity of $\nu_{1a}$ and $\nu_{1c}$ in lycopene for a range of resonance conditions (Raman pump varied between 540 & 610 nm; Fig. **4C**). The observed changes in the $\nu_{1a}/\nu_{1c}$ intensity ratio suggests that they arise from distinct excited states. On the other hand, the global fitting clearly showed that they decay synchronously - indicating either strong coupling or rapid equilibration between the two underlying states. Consequently, $\nu_{1c}$ does not exhibit independent kinetics despite being an apparently independent state, and so we have to distinguish it by other means.

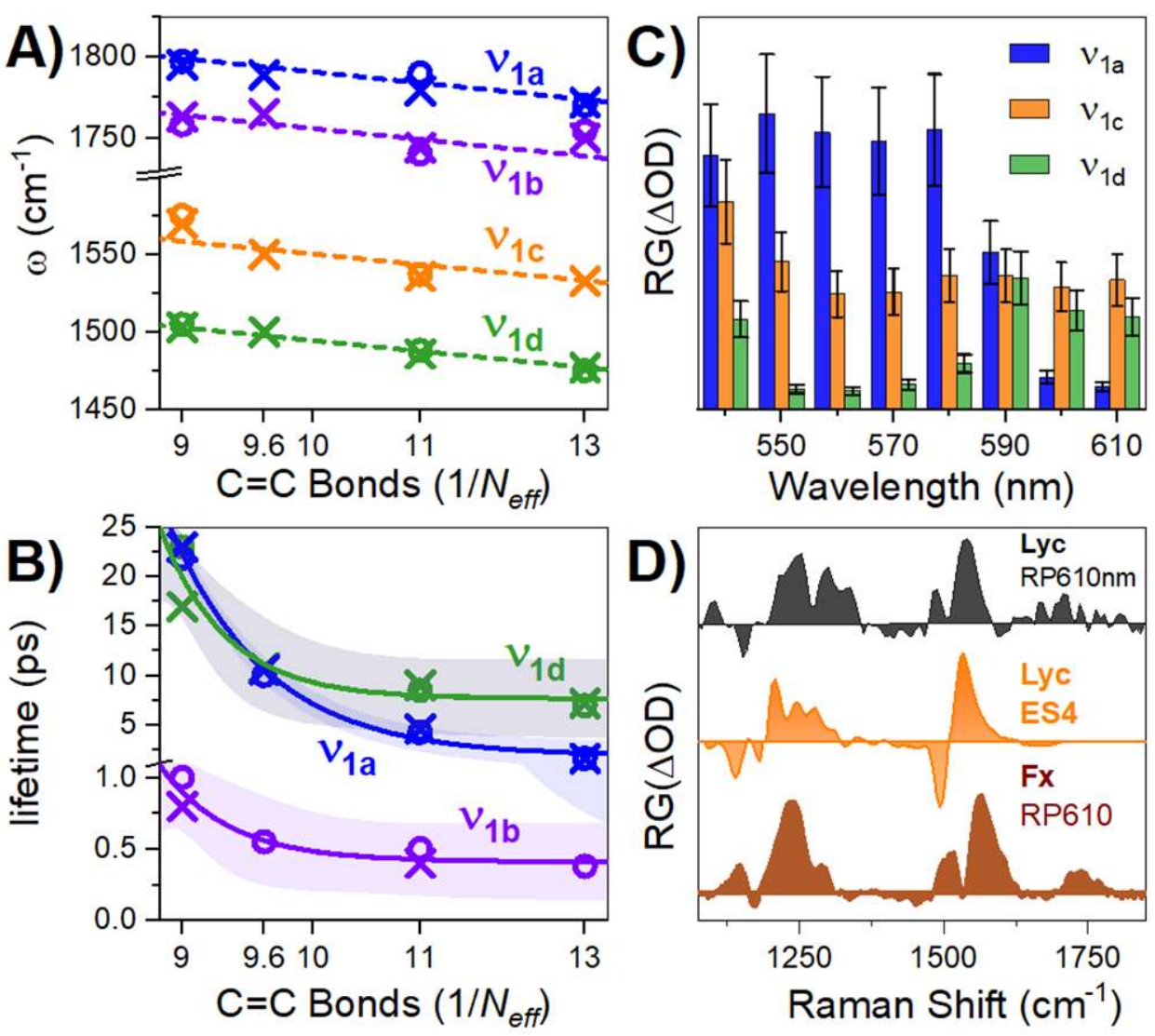

**Figure 4**. **(A, B)** Peak positions and kinetics (respectively) of $\nu_{1(a\text{-}d)}$ modes of carotenoids in n-hexane (○) and THF (×). The peak is taken as the maximum intensity of the gated Raman profile. Exponential fits of the kinetics, following the energy-gap law, are given by solid lines, and shading represents the 90 % confidence region. Note x-axes are reciprocal (1/x). **(C)** RP-dependent changes in relative intensity of $\nu_{1a}$, $\nu_{1d}$ and $\nu_{1c}$ modes for lycopene in THF. **(D)** Time-gated Raman spectra for lycopene in THF (black; actinic pump 510 nm, RP 610 nm) and fucoxanthin in methanol (maroon; actinic pump 485 nm, RP 610 nm). ES4 (orange) is reproduced from Fig. **3B** for comparison.

Modes in the 1530-1550 cm$^{-1}$ region correspond to a very large upshift in the $\nu_1$ band, only observed in fucoxanthin and echinenone when promoting the intramolecular charge-transfer (ICT) state [8, 27-28]. In figure **4D**, we compare fucoxanthin in methanol to lycopene in THF at Raman pump 610 nm, along with the ES4 SAS from Fig. 3B (lycopene in THF; RP 540 nm). At this wavelength, on the red side of the $S_1$ electronic transition, fucoxanthin exhibits a 1560 cm$^{-1}$ band unambiguously assigned to the fucoxanthin ICT state [8, 27-28]. Lycopene displays an equivalent band at 1542 cm$^{-1}$ that decays with $\tau$ = 4.2 ps (Table 1). Neither carotenoid displays any significant $S_1$ signature above 1700 cm$^{-1}$ at this RP excitation, whereas these appear for fucoxanthin around 1740 cm$^{-1}$ when the Raman pump is moved to 550 nm (Supplementary Fig. S4-5). In keto carotenoids like fucoxanthin, increasing the solvent polarity stabilises ICT, promotes $S_1$ →ICT population flow, and shortens the $S_1$ lifetime [29-31]. We did not observe any significant polarity dependence of the ICT state for the non-carbonyl carotenoids measured here (see Supplementary Information Fig. S4). The observation of an ICT-like feature in a symmetric carotenoid is unexpected. We therefore performed concentration-dependent measurements to test whether excitonic interactions could account for its appearance. As shown in Figure S14 (Supporting Information), the intensity ratio of the vibrational band assigned to the ICT-like feature relative to the band assigned to $S_1$ remains unchanged across the concentration range, arguing against an excitonic origin.

The target model provides a satisfactory global fit to the decay of the $\nu_{1b}$ band (ES2) and the concomitant rise of $\nu_{1a}$ (ES3). However, the SAS associated to $\nu_{1b}$ is highly asymmetrical which would not be consistent with an independent excited-state population, where a symmetric band would rise and decay following the population kinetics. Hence, this calls at least for a revision of the target model and its limitations. Because the target model does not include vibrational cooling or spectral relaxation, we therefore complement it with a more direct analysis of the raw data, using only baseline removal to subtract the underlying TA contribution. Figure 5 presents room-temperature FSRRS time–frequency maps in the 1650–1900 cm$^{-1}$ region for the three linear carotenoids, under resonance conditions that enhance the $\nu_{1b}$ band. The side panels show time-gated spectra at selected delays. None of the datasets displays an isosbestic point, which would be expected for a clean interconversion between two species. Instead, the band maximum shifts progressively to higher frequency while the feature narrows. This behaviour argues against $\nu_{1b}$ reflecting an independent excited state and is more consistent with vibrational relaxation within the same electronic manifold.

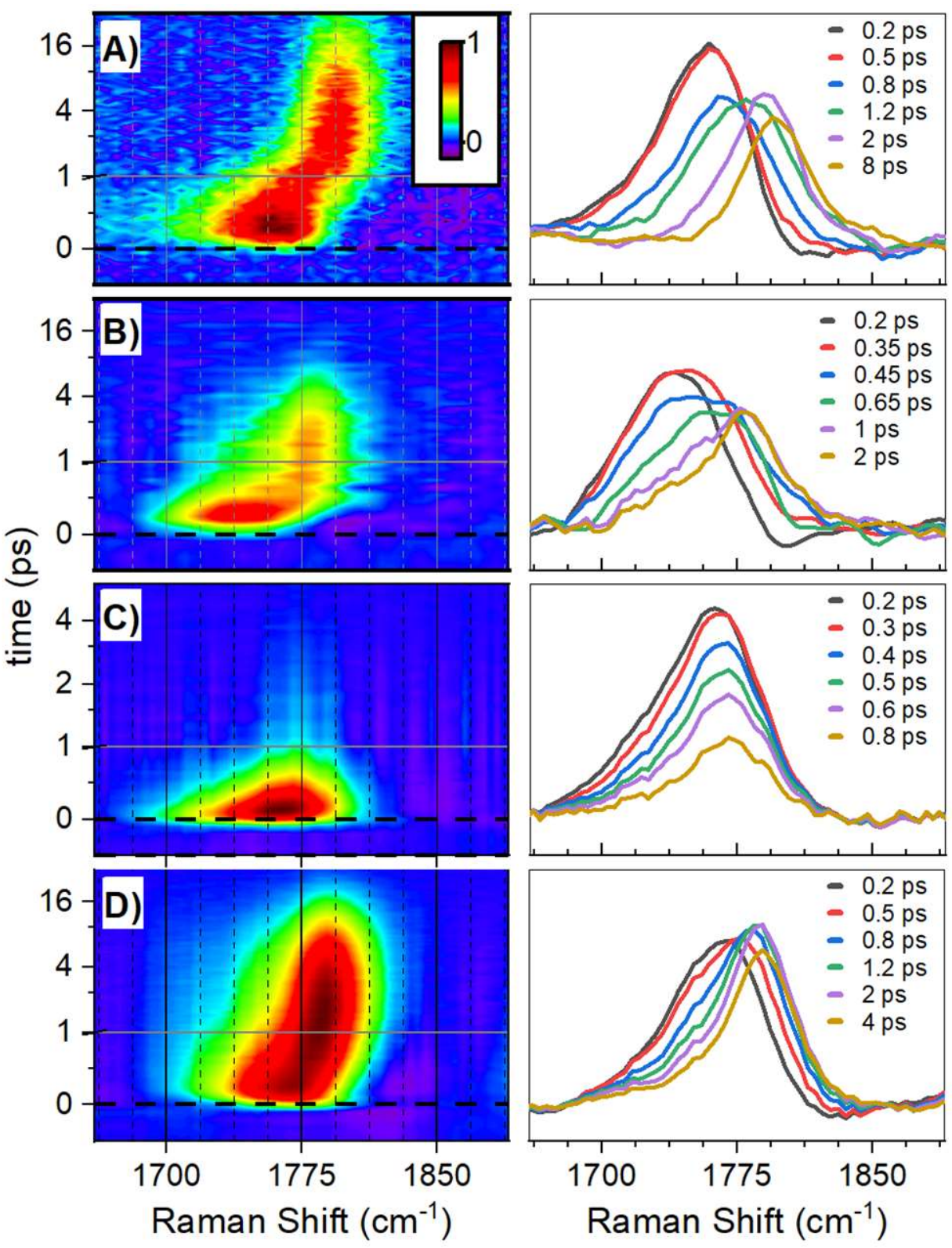

**Figure 5**. Femtosecond stimulated resonant Raman spectroscopy of carotenoids at room temperature in the 1650–1900 cm$^{-1}$ region. Colour maps (left panels) show Raman intensity as a function of Raman shift (cm$^{-1}$) and pump–probe delay, displayed on a linear time axis up to 1 ps and on a logarithmic axis at longer delays. Data are shown for (A) neurosporene in n-hexane (actinic pump 475 nm; Raman pump 530 nm), (B) lycopene in THF (actinic pump 510 nm; Raman pump 590 nm), (C) spirilloxanthin in THF (actinic pump 540 nm; Raman pump 600 nm), and (D) β-carotene in THF (actinic pump 485 nm; Raman pump 570 nm). For each carotenoid, the right panel presents time-gated spectra extracted at the indicated delays.

**Nature of electronic excited states**.

The nature of dark states, as well as the structure of potential energy surfaces (PES) at Franck Condon excitations and their conical intersections remains under debate. For strongly diabatic PES, bond length alternation (BLA) coordinate has been proposed to capture the key couplings [32], but the predicted dynamics fall below our time resolution and do not account for signals at t > 0.15 ps. Furthermore, BLA alone is insufficient to describe full conformational dynamics and symmetry-dependent vibrational couplings, motivating the revised state assignments developed below. In the following discussion we adopt the conventional language of distinct electronic states ($S_1$, hot-$S_1$, ICT, S*) as a tractable framework for organising the observations. An alternative description, in terms of a dynamically evolving, strongly coupled $B_u/A_g$ manifold where geometry-dependent BLA changes drive interconversion between regions of mixed $B_u$ and $A_g$ character, could be compatible with our data and cannot at present be excluded. The state labels used below could therefore be read as identifying regions of the excited-state manifold with characteristic vibrational signatures, rather than as fully separable electronic species.

**$S_2$ State**. The first component in the global analysis appears convoluted with the laser pulse (IRF 0.15 ps), and cannot be resolved by our system. However, it has been characterised recently showing a wide positive feature on the high energy side of the ground state $\nu_1$ mode, assigned to C=C modes of the $S_2$ state ($1^1B_u^+$) [11]. The change in symmetry between ground state $S_0$ ($1^1A_g^-$) and $S_2$ ($1^1B_u^+$) does not favour or induce vibronic coupling, so the vibrational mode is *circa* 1600 cm$^{-1}$. This state decays in less than 150 fs, giving rise to states associated with modes $\nu_{1b}$ and $\nu_{1d}$, respectively.

**$S_1$ State**. The $S_1$ state is associated with the $\nu_{1a}$ band (region 1770–1800 cm$^{-1}$) [4]. Vibrational modes in regions $\nu_2$ and $\nu_3$ for neurosporene, extracted by target analysis, contribute at 1045 and 1282±50 cm$^{-1}$ (wide envelope), respectively. Equivalent modes for lycopene and spirilloxanthin are observed at 1024, 1060, 1208, 1246 & 1275 cm$^{-1}$, and 1000, 1030, 1222, 1254 & 1288 cm$^{-1}$, respectively. Whereas these $\nu_3$ & $\nu_2$ regions mirror the ground-state Raman spectra quite closely (see supporting Information Fig. S1), the $\nu_{1a}$ mode exhibits an extraordinarily high upshift ($\nu_1 \approx 1520$ cm$^{-1}$ in the ground state). This large frequency shift has been shown to arise from a strong vibronic coupling between the $S_0$ ($1^1A_g^-$) and $S_1$ ($2^1A_g^-$) states, through a g-type C=C stretching symmetric vibration [33-35]. Therefore, the presence of this large upshift for the C=C stretching mode is consistent with the view that the $S_1$ electronic state has an $A_g$ character that can induce vibronic coupling with the ground state $1^1A_g^-$.

**Vibrationally hot $S_1$**. The $\nu_{1b}$ mode (region 1730–1760 cm$^{-1}$), more easily observed for RP energies to the red of $S_1$ , appears in <150 fs and evolves into $S_1$ in < 1 ps (the precise lifetime varies according to $N_{eff}$, Fig. 4B). Regions $\nu_2$ and $\nu_3$ can be extracted for lycopene (1024 & 1060, and 1208, 1246 & 1275 cm$^{-1}$) and spirilloxanthin (1000 & 1030, and 1222, 1254 & 1288 cm$^{-1}$). These modes are exactly the same as observed for $S_1$ above, whereas the $\nu_1$ mode is downshifted by 30-40 cm$^{-1}$. This band has previously been associated either with a vibrationally hot $S_1$ or with $S_x$ [4-10, 14, 36-39]. When associated to Sx excited state, it has generally been assigned to $1^1B_u$. However, the high frequency observed for $\nu_{1b}$, and the similarity of the $\nu_2$ and $\nu_3$ frequencies with those for the $S_1$ state, call for a revision. These frequency values suggest strong vibronic coupling to another $A_g$ state, similar to that described above between $S_0$ ($1^1A_g^-$) and $S_1$ ($2^1A_g^-$). This $A_g$ state should be close in energy to the $1^1B_u^-$ state, which would correspond to the $3^1A_g^-$ state described in the literature[40-41] but it should be higher in energy than $1^1B_u^-$ state. On the other hand, the strong vibronic coupling is a first argument to assign this state to a hot vibronic $S_1$. Then, as shown in figure 5, the evolution of the vibrational mode from $\nu_{1b}$ into $\nu_{1a}$ ($S_1$) is not readily consistent with the presence of an independent state, and is most simply described as a vibrationally hot $S_1$.

Such behaviour is consistent with vibrational relaxation on an anharmonic potential. The actinic excitation prepares population in $S_2$ that transfers to $S_1$ on a timescale shorter than 150 fs, i.e., within the spectral bandwidth of the excitation pulse (∼100 cm$^{-1}$). This impulsive transfer results in the population of a distribution of vibrational levels in the $S_1$ state. Due to the anharmonic nature of the potential, the vibrational level spacing varies as energy increases. Higher vibrational states are characterised by smaller level spacings, whereas larger spacings are encountered closer to the bottom of the potential. As the system relaxes within the $S_1$ manifold, progressively lower vibrational states are populated, leading to the observed shift of the Raman frequency[12-13]. At the same time, the initial population of multiple vibrational levels produces a broader spectral distribution, which narrows as vibrational cooling funnels population toward the lowest vibrational state. The concomitant decrease in FWHM therefore reflects the collapse of the initially broad vibrational distribution during the relaxation process [37] .

**ICT State**. All carotenoids studied in this work exhibit a $\nu_{1c}$ band in the 1530–1580 cm$^{-1}$ window, decaying at the same rate as $\nu_{1a}$ ($S_1$), indicating strong coupling or equilibration within < 2 ps. The ∼1550 cm$^{-1}$ feature overlaps a broad early-time $S_2$-associated band, preventing us from determining if this state forms directly from $S_2$ [28] - but if a short-lived intermediate state exists between them then it is neither $S_1$ or hot-$S_1$, since the latter states decay slower than the rise time of ICT. Nonetheless, $\nu_{1a}$ and $\nu_{1c}$ reach quasi-equilibrium much faster than the decay time of $S_1$ (2-20 ps). Previous work has shown that ICT and $S_1$ can be strongly coupled, approaching rapid energy equilibration ("dynamic equilibrium") on sub-ps timescales [28, 42]. While the presence of an ICT state in a fully symmetrical carotenoid molecule is formally forbidden, our experiments are performed in solution at room temperature, conditions where these molecules will display transient deviations from C2h symmetry. Considering that resonance conditions may pick up even weakly-populated species, it may be that these observed ICT states arise from molecules in conformations distorted enough to break their formal symmetry, and allow an excited state with a charge-transfer character to appear. Although ICT is classically associated with carbonyl substitution, there is no formal prohibition against ICT character in neutral carotenoids, provided their symmetry can be dynamically broken. Indeed, even non-carbonyl carotenoids show a small but reproducible red-wing broadening of the absorption profile with increasing solvent polarity, consistent with emergent ICT character [43], and a minor ICT contribution has indeed been proposed for neutral carotenoids recently [11]. It is also noticeable that despite the enhancement provided via resonance the vibrational modes associated to ICT are more prominent in long carotenoids, which generally exhibit greater conformational flexibility. These observations support a tentative ICT-like assignment for $\nu_{1c}$ in neutral carotenoids. Clearly more studies are necessary to assess the precise origin of this state, as well as whether it is generally present or only in a subpopulation of carotenoids.

**S* State**. The $\nu_{1d}$ mode (region 1475-1505 cm$^{-1}$) has already been associated with S* [4]. It rises within < 150 fs—ruling out $S_1$ or $S_x$ as precursors—and decays independently from these states. Thus, S* is formed either directly from $S_2$ or *via* an even shorter-lived intermediate that we cannot resolve.

The full vibrational signatures can be disentangled, showing the S* vibrational fingerprints across $\nu_1$ (1475-1505 cm$^{-1}$), $\nu_2$ (1030 – 1140 cm$^{-1}$), and $\nu_3$ (~1010 cm$^{-1}$). These bands coincide with carotenoid triplet signatures observed on microsecond timescales [44-45] and by power-induced Raman [46]. Triplet formation is accompanied by a ~25 cm$^{-1}$ downshift of the symmetric C=C stretching mode, reflecting a bond-order inversion $S_0 \rightarrow T_1$[47].

The $\nu_{1d}$ downshift alone is not conclusive proof of triplet character. In the literature, the “hot-ground-state interpretation” [3-4] attributes $\nu_{1d}$ to the anharmonically-shifted $\nu_1$ of vibrationally hot $S_0$ formed by ultrafast internal conversion, and can account for both the ~25 cm$^{-1}$ frequency shift and the picosecond decay timescale (vibrational cooling). McCamant et al.[48] measured anti-Stokes resonance Raman spectra of hot $S_0$ for β-carotene, and established the following constraints: the hot-$S_0$ $\nu_1$ (C=C) shifts by ~14 cm$^{-1}$ from 1523 to ~1509 cm$^{-1}$ and cools in ~12 ps; the hot-$S_0$ $\nu_2$ (C–C) shifts by ~12 cm$^{-1}$ from 1157 to ~1145 cm$^{-1}$ and cools in ~5 ps; the hot-$S_0$ $\nu_3$ (C–$H_3$, ~1005 cm$^{-1}$) shows negligible frequency shift and cools in ~2 ps. Critically, hot $S_0$ produces no new spectral features at anomalous frequencies; the same modes appear at the same positions with enhanced anti-Stokes intensity. Each mode cools at a different rate, reflecting mode-specific energy dissipation pathways. The $\nu_2$ and $\nu_3$ regions provide far more compelling evidence than $\nu_1$ for distinguishing between these interpretations. This spectral window was not accessible to previous FSRS studies employing off-resonant near-IR Raman pumps, where solvent interference and lower signal-to-noise obscured the C–C stretching modes. Our use of resonant visible Raman pump wavelengths provides the signal-to-noise necessary to resolve the $\nu_2$ region, and gated spectra at representative time delays are now shown in the Supporting Information (Fig. S3).

For β-carotene, the $\nu_{1d}$ mode at 1496 cm$^{-1}$ with a lifetime of 10.8 ps could be considered broadly consistent with an anharmonically-shifted hot-$S_0$ $\nu_1$, even though the downshift of ~27 cm$^{-1}$ already strains this assignment. Considering $\nu_2$, the measured anharmonic ceiling by McCamant et al. [48] places the lowest possible hot-$S_0$ $\nu_2$ frequency at ~1145 cm$^{-1}$. Our data exhibits $\nu_2$ features centred at ~1130 cm$^{-1}$, a shift of ~30 cm$^{-1}$ from the $S_0$ $\nu_2$ centre, exceeding the measured ceiling by ~15 cm$^{-1}$. This feature decays with a lifetime of 10.8 ps, matching the $\nu_{1d}$ decay rather than the 5 ps cooling time established for the hot-$S_0$ $\nu_2$ mode. In the $\nu_3$ region, the peak position does not change significantly, but it decays together with $\nu_{1d}$ and $\nu_2$ at ~10.8 ps, in contrast to the 2 ps cooling time of the hot-$S_0$ $\nu_3$ mode. The simultaneous decay of all three modes at the same rate is the hallmark of a single electronic species depopulating, not of mode-specific vibrational cooling in the ground state, where each mode cools independently at its own rate. A comparison of the frequency shifts and lifetimes for β-carotene, lycopene and spirilloxanthin (supporting information, Fig. S19) confirms this pattern: all three modes show shifts of 20–30 cm$^{-1}$ from the ground-state values and decay simultaneously, inconsistent with hot-$S_0$ cooling in every case.

The $\nu_1$ and $\nu_2$ pattern, with features at 1030–1140 cm$^{-1}$, is consistent with the established triplet Raman signatures of carotenoids [44-47] . Tavan and Schulten[49] calculated that the triplet state locally inverts bond-length alternation, causing formerly single C–C bonds to acquire double-bond character and producing vibrational modes downshifted by ~30 cm$^{-1}$ to frequencies inaccessible by an anharmonically-shifted $S_0$. This prediction was confirmed by Ho et al. via DFT calculations [47]. The absence of efficient energy transfer between S* and $S_1$ lends further support to this assignment - if S* has ungerade (u) symmetry, spin–orbit coupling will not mix it effectively with the $S_1$ ($2^1A_g^-$) state [50]. While the vibrational signatures for S* indicate a triplet state, the question of its short lifetime remains. However, this can be explained by the existence of triplet-pair amplitudes in singlet wavefunctions, which opens an excited-state manifold consisting of doubly-excited configurations, $^1$(TT)*, comprising a distribution of triplet separations, geometries, and internal energies. This manifold broadens the stimulated-absorption band and undergoes ultrafast relaxation in the ps range, as for the $2^1Ag^-$ state [51-52]. Tavan and Schulten predicted that the probability of a covalent singlet dissociating into its triplet constituents is strongly enhanced by distortions of the polyene chain [49], which is a feature observed experimentally for long carotenoids or produced by distortion of the carotenoid in some photosynthetic proteins [26, 53]. Theoretical calculations applied to long polyenes predicted that several covalent excitations (e.g. $1^1B_u^-$ , $3^1A_g^-$ , etc.) will be situated below the ionic $1^1B_u^+$ ($S_2$) state [49], being “–” states with the character of multiple triplet excitations. A singlet covalent excitation (such as the $S_1$ state of a polyene) can be viewed as two spin-correlated triplets [$^3B_u \otimes {}^3B_u$] coupled into an overall singlet state [49, 54].

The triplet character of S* [$^3B_u \otimes {}^3B_u$] is supported by the convergent vibrational signatures discussed above: the simultaneous decay of $\nu_{1d}$, $\nu_2$ and $\nu_3$ at a common rate, the position of the $\nu_2$ band below the hot-$S_0$ ceiling, and the close correspondence of the mode pattern with known carotenoid triplet signatures. Complementary measurements, for example magnetic-field-dependent experiments, would provide further constraints on this assignment. Symmetry considerations also suggest the existence of an intermediate between $1^1B_u^+$ and S* - plausibly this state could be associated with the previously proposed $1B_u^-$ state, based on its close energetic position and its nature as a [$^3B_u \otimes {}^3A_g$] correlated pair. We did not detect such an intermediate, probably because it forms and decays below our temporal resolution (< 150 fs), but its energetic proximity and symmetry make it a reasonable candidate. In our data, S* shows distinctive triplet-like vibrational features, whereas no separate $A_g$-type signal could be isolated within the IRF. Consistent with this picture, Polívka and co-workers reported sub-0.5-ps FSRS features in the $A_g$ region (~1700 cm$^{-1}$) that relax through $B_u$-type bands (1100–1200 and ~1500 cm$^{-1}$) in lycopene and spirilloxanthin - behaviour compatible with a [$^3B_u \otimes {}^3A_g$] intermediate involved in S* formation [11]. Targeted experiments will be required to confirm the identity of [$^3B_u \otimes {}^3A_g$].

## Conclusions

In conclusion, tunable femtosecond stimulated resonance Raman spectroscopy yields state-selective vibrational fin-gerprints for linear carotenoids, resolving four reproducible $\nu_1$ mode contributions that map onto distinct excited states: $\nu_{1a} \rightarrow S_1$, $\nu_{1b} \rightarrow$ hot vibrational $S_1$, $\nu_{1c} \rightarrow$ ICT, and $\nu_{1d} \rightarrow$ S*. The full spectral region resolved here exhibits vibrational features that allow us to assign the nature and symmetry of the hot vibrational $S_1$, ICT, and $S^*/^1(TT)^*$ [$^3B_u \otimes {}^3B_u$]. The presence of the ICT state, previously associated primarily with keto-containing carotenoids, may have broader relevance to other carotenoid types. Furthermore, we have resolved vibrational signatures of S* consistent with an entangled triplet-pair character, distinct from those of the well-studied $S_1$ state. The distinct spectral signatures we have identified for each carotenoid dark state should prove invaluable for assessing their contribution to energy transfer and photoprotective processes in photosynthetic proteins.

## ASSOCIATED CONTENT

The Supporting Information is available free of charge at http://pubs.acs.org.

Materials and methods; ground-state resonance Raman spectra of all carotenoids (Fig. S1); femtosecond transient absorption maps, time-gated spectra, and kinetics for all carotenoids (Fig. S2); time-gated FSRRS spectra of the linear carotenoids (Fig. S3); FSRRS time–spectral maps at all Raman pump wavelengths and solvents (Figs. S4–S6); selection of the kinetic model, with sequential and branched target analyses, species-associated spectra, and fits (Figs. S7–S12); Raman-pump-frequency dependence of the relative band intensities (Fig. S13); concentration dependence of the FSRRS spectra (Fig. S14); Raman-pump actinic-effect control and Raman-pump fluence dependence (Figs. S15–S18); time-gated Raman spectra supporting the simultaneous multi-mode decay of S* (Fig. S19)

## AUTHOR INFORMATION

### Corresponding Author

* *Juan Jose Romero.* Email : juan-jose.romero@i2bc.paris-saclay.fr
* *Bruno Robert.* Email : bruno.robert@cea.fr
* *Manuel J. Llansola-Portoles.* Email *:* manuel.llansola@cnrs.fr

### Funding Sources

Agence Nationale de la Recherche (ANR) SINGLETFISSION grant ANR-23-CE29-0007 (ML).
Agence Nationale de la Recherche (ANR) FISCIENCY grant ANR-23-CE50-009 (ML).
France 2030 PEPR LUMA program SYNFLUX-LUMICALS grant ANR-23-EXLU-0001(ML).
France 2030 PEPR LUMA program ULTRAFAST platform grant ANR-22-EXLU-0002 (BR).
French Infrastructure for Integrated Structural Biology (FRISBI) grant ANR-10-INSB-05 (I2BC Biophysics platform)(BR, AP).

---

## Table of Contents artwork

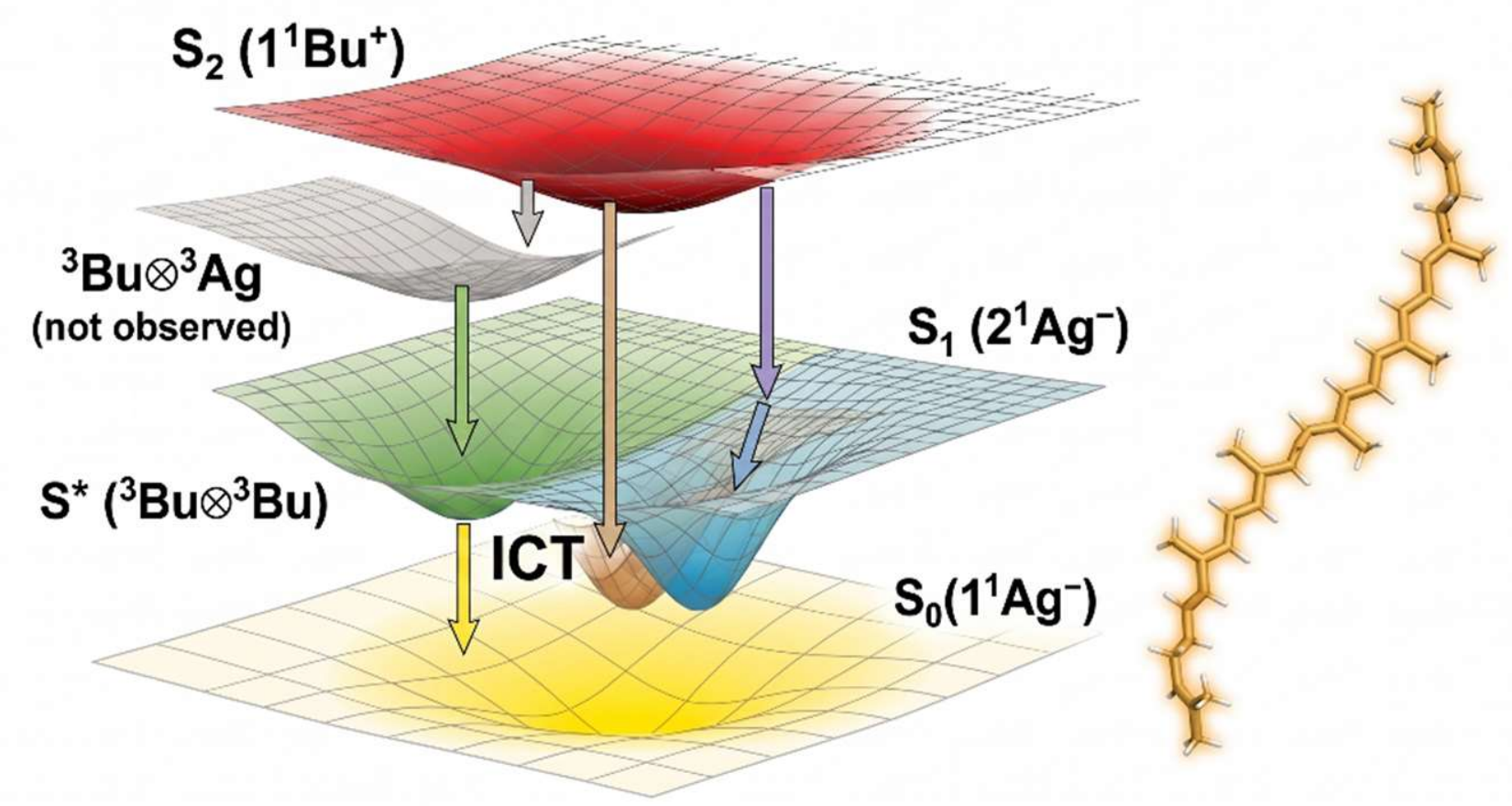

---

# Supplementary Information

## Reassessing carotenoid photophysics: shedding light on dark states

Roxanne Bercy, Viola D'mello, Andrew Gall, Cristian Ilioaia, Andrew A. Pascal, Juan Jose Romero*, Bruno Robert*, Manuel J. Llansola-Portoles*

Institute for Integrative Biology of the Cell, CEA, CNRS, Université Paris Saclay, CEA Saclay 91191 Gif sur Yvette Cedex, France.

Corresponding authors:

Juan J. Romero. Email: juan-jose.romero@i2bc.paris-saclay.fr

Bruno Robert. Email : bruno.robert@cea.fr

Manuel J. Llansola-Portoles. Email : manuel.llansola@cnrs.fr

## Materials and Methods

Carotenoid purification**.** Lycopene was extracted from tomato paste by stirring a biphasic dichloromethane (DCM)/water mixture (80:20, v/v) for 1 hour. The organic phase was then separated, concentrated under reduced pressure, and purified by silica-gel column chromatography (hexane, followed by hexane/acetone 95/5 up to hexane/acetone 90/10 as gradient) (*1*). Spirilloxanthin was extracted from *Rhodospirillum rubrum* S1 membranes (in 20 mM MES, 100 mM KCl, pH 6.8 buffer)(*2*) in DCM. The organic phase was obtained as described above then separated by thin-layer chromatography; the spirilloxanthin band was identified, scraped, and eluted from the silica by filtration. Neurosporene was purchased from Carotenature, while β-carotene (96 %) and all other chemicals and solvents were obtained from Sigma-Aldrich; all were used without further purification.

Steady State Absorption and Raman measurements. Absorption spectra were measured using a Varian Cary E5 scanning spectrophotometer, using a square cell with a 1 cm path length. Resonance Raman spectra were recorded at room temperature with excitations obtained from a Coherent $Ar^+$ (Sabre) laser. Output laser powers of 10–100 mW were attenuated to < 5 mW at the sample. Scattered light was collected at 90° to the incident light, and focused into a Jobin-Yvon U1000 double-grating spectrometer (1800 grooves/mm) equipped with a red-sensitive, back-illuminated, LN2-cooled CCD camera. Sample stability and integrity were assessed based on the stability of the Raman signal.

Time-resolved femtosecond transient absorption (fsTA) and femtosecond stimulated resonance Raman spectroscopy (FSRRS). Samples were transferred into a 1 mm path-length fused-silica cuvette, adjusted to an OD of 0.6 - 0.8 (for fsTA) or 0.1 – 2.0 (for FSRRS) at the absorption maximum, and deoxygenated by purging with $N_2$ for 30 min. Sample integrity was monitored by recording the steady-state absorption spectrum before and after each run, and fresh aliquots were used whenever the absorbance changed by more than 10 %. A PHAROS Yb:KGW femtosecond laser (Light Conversion; 1030 nm, 120 fs, 10 kHz, 10 W) was split into three beams to generate the white-light probe (WL), actinic pump (AP) and Raman pump (RP), as described elsewhere (*3*). White-light probe (WL): A fraction of the 1030 nm fundamental was attenuated by a neutral-density filter and focused into a 13 mm sapphire plate to generate a broadband continuum (~480–1100 nm). Actinic pump (AP): A separate portion of the fundamental fed an ORPHEUS HE optical parametric amplifier, yielding wavelength-tunable pulses (∼150 fs FWHM; ∼100 $cm^{-1}$ bandwidth; pump fluence 150 μJ $cm^{-2}$). Raman pump (RP): The remaining beam seeded a second-harmonic band compressor (SHBC), whose output drove an ORPHEUS PS OPA and LYRA difference-frequency stage to deliver a narrowband Raman pump (∼5 $cm^{-1}$ resolution; ∼3 ps FWHM; pump fluence 400 - 600 μJ $cm^{-2}$). All three beams were directed through independent delay stages into a HARPIA-TA spectrometer. AP and RP beams were chopped at 1 and 0.5 Hz, respectively; the transmitted WL was spatially filtered, collimated, and dispersed by an Andor Kymera 193i spectrograph (Oxford Instruments) onto a 256-pixel Hamamatsu S8380 diode array (200–1100 nm). The femtosecond probe was temporally matched with the leading edge of the Raman pump, in order to minimise any actinic contribution from the Raman pump itself. At each pump–probe delay, the transient absorption contribution was first subtracted; any residual baseline was then removed by fitting a 9th-order polynomial to spectral regions free of Raman bands and subtracting the result.

*Broadband TA*: 300 g/mm grating (800 nm blaze); $\Delta A(t,\lambda)$ obtained from the integrated intensity difference between 2000 pumped and 2000 unpumped WL pulses per delay.

*FSRRS:* 1200 g/mm grating (600 nm blaze); spectra computed over 5000 WL shots using a four-state chopping scheme: I(RP on, AP on), I(RP off, AP on), I(RP on, AP off), and I(RP off, AP off)(*4*). Grating calibration was carried out using carotenoid resonance Raman peaks (Fig S10). Global data analysis was performed using CARPETVIEW software (Light Conversion)(*5*).

**Supplementary Figures**

Resonance Raman

Continuous-wave resonance Raman spectra, probing ground state vibrational structure, display four main groups of bands for carotenoids, denoted $\nu_1$ to $\nu_4$, which can be satisfactorily modelled by Density Functional Theory calculations (*6-9*). The highest frequency $\nu_1$ mode above 1500 $cm^{-1}$ arises from stretching vibrations of C=C double bonds. Its position depends on the length of the $\pi$-electron conjugated chain and on the molecular configuration of the carotenoid (*10, 11*). The $\nu_2$ envelope around 1160 $cm^{-1}$ contains contributions from stretching vibrations of C-C single bonds coupled with C-H in-plane bending modes, and this region is a fingerprint for the assignment of carotenoid isomerization states (*cis*/*trans*) (*12-14*). The $\nu_3$ band at 1000 $cm^{-1}$ arises from in-plane rocking vibrations of the methyl groups attached to the conjugated chain, coupled with in-plane bending modes of the adjacent C-H's (*15*). It was reported to be a fingerprint of the configuration of conjugated end-cycles (*16, 17*), a hypothesis which is supported by theoretical modelling (*9*). Finally, the $\nu_4$ band around 960 $cm^{-1}$ arises from C-H out-of-plane wagging motions coupled with C=C torsional modes (out-of-plane twists of the carbon backbone) (*10*). When the carotenoid conjugated system is planar, these out-of-plane modes are not coupled with the electronic transition (which is oriented along the plane), and so they exhibit little resonance enhancement. However, distortions around C-C single bonds increase the coupling of (some of) these modes with the electronic transition, resulting in an increase in their intensity (*10, 18*).

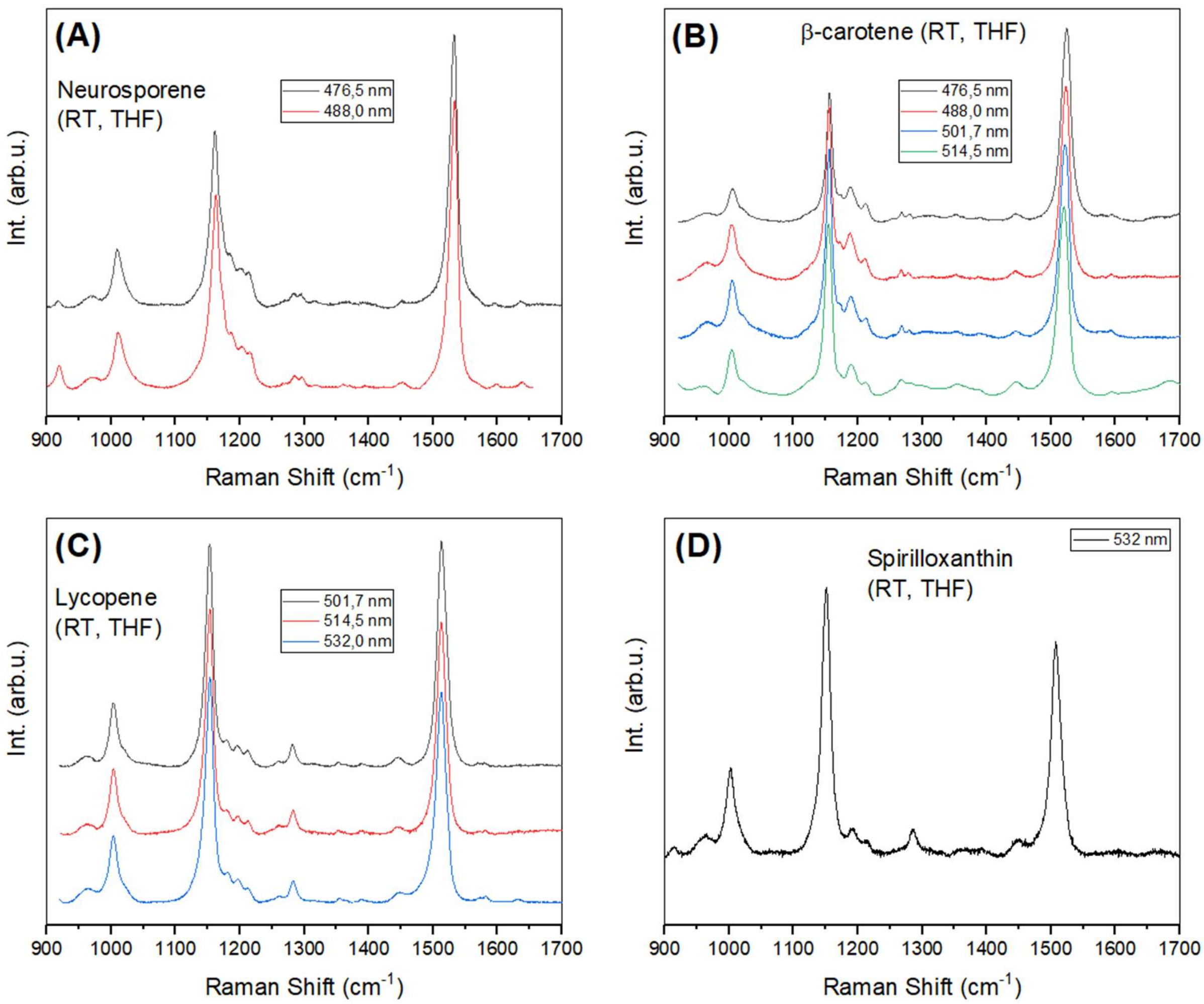

**Figure S1 |** Resonance Raman spectra at room temperature in THF, excited near the 0-0 electronic transition, for **(A)** neurosporene ($N_{eff} \approx 9$), **(B)** β-carotene ($N_{eff} \approx 9.6$), **(C)** lycopene ($N_{eff} \approx 11$), **(D)** Spirilloxanthin ($N_{eff} \approx 13$). Excitation wavelengths are indicated in the legends of each panel.

## Femtosecond Transient Absorption

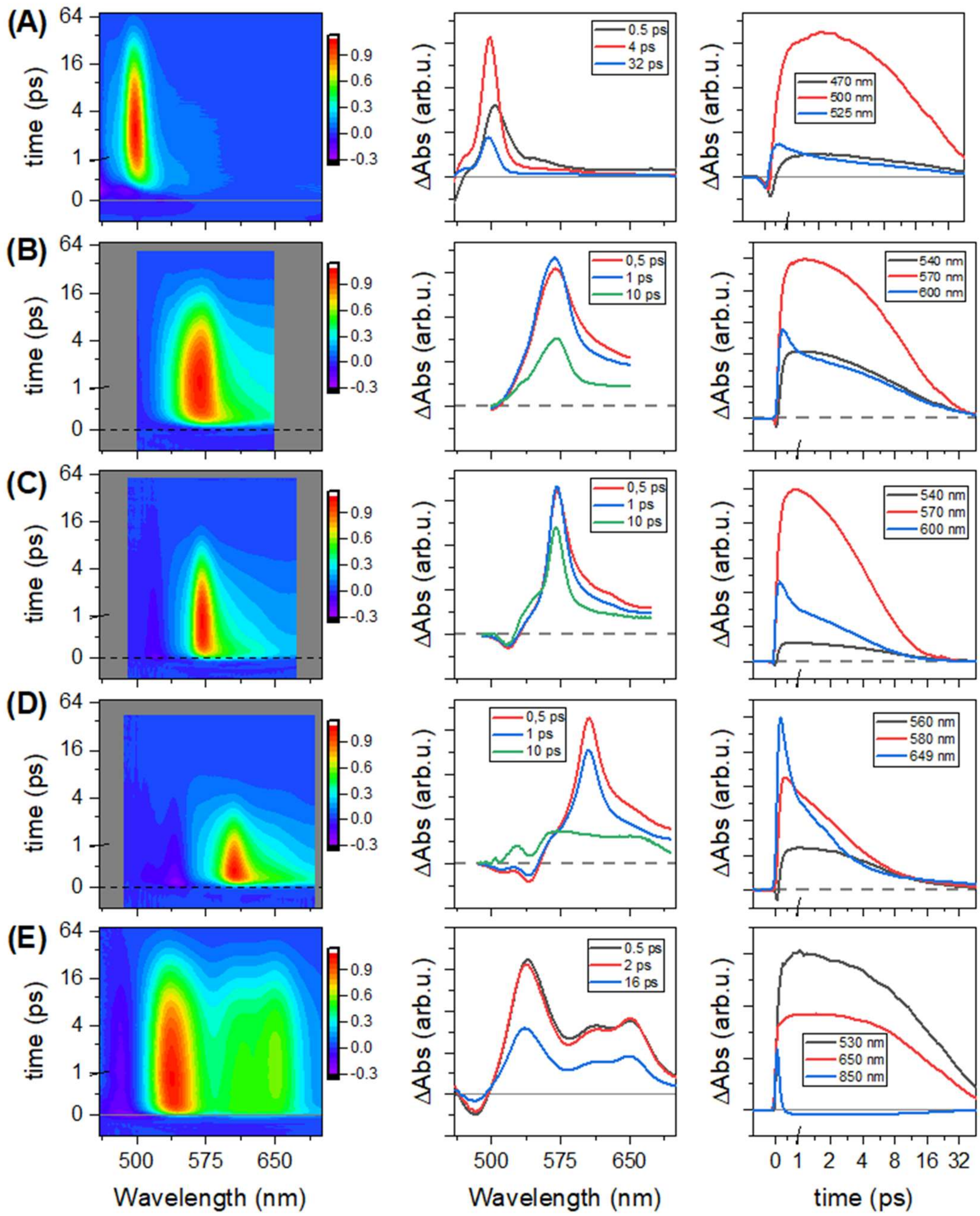

**Figure S2|** Femtosecond transient absorption at 298 K of (A) neurosporene in n-hexane, excitation 445 nm; (B) β-carotene in THF, excitation 435 nm; (C) lycopene in THF, excitation 450 nm; (D) spirilloxanthin in THF, excitation 470 nm; (E) fucoxanthin in methanol, excitation 440 nm. Left panels show the color-coded 2D map of the differential absorbance (ΔAbs) as a function of wavelength (nm) and pump–probe delay (–1 to 50 ps; linear scale to 1 ps, logarithmic thereafter). Central panels show the time-gated transient absorption spectra (ΔAbs) extracted from the 2D map at selected pump–probe delays. Right panels show the representative kinetic traces of ΔAbs at the ground-state bleach and excited-state absorption bands, plotted on a linear time axis to 1 ps and logarithmic thereafter.

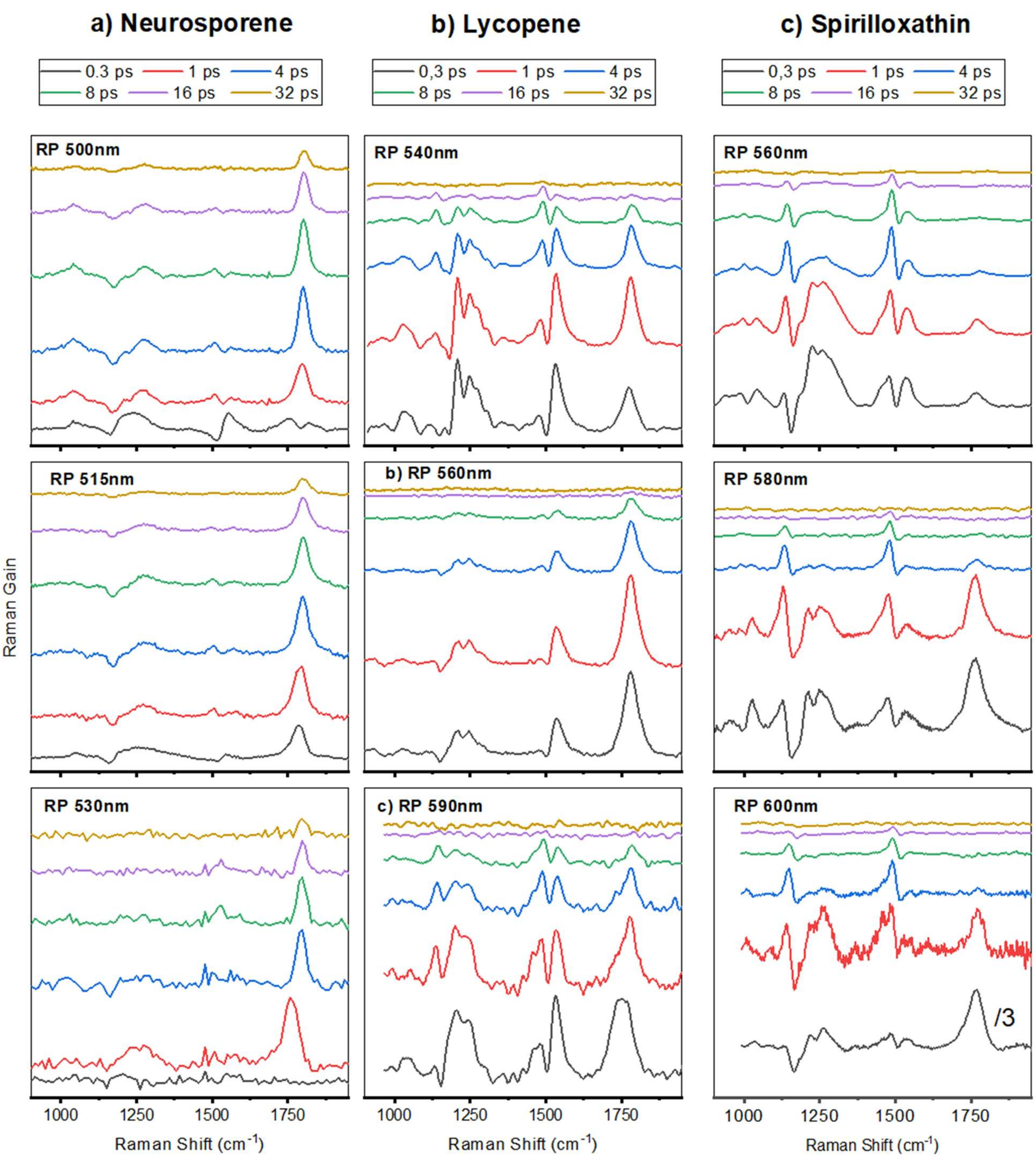

**Figure S3** | Time-gated femtosecond stimulated resonance Raman (FSRRS) spectra of carotenoids at room temperature. (a) Neurosporene in n-hexane, excited at 475 nm, with Raman pump wavelengths of 500, 515, and 530 nm. (b) Lycopene in THF, excited at 510 nm, with Raman pump wavelengths of 540, 560, and 590 nm. (c) Spirilloxanthin in THF, excited at 540 nm, with Raman pump wavelengths of 560, 580, and 600 nm.

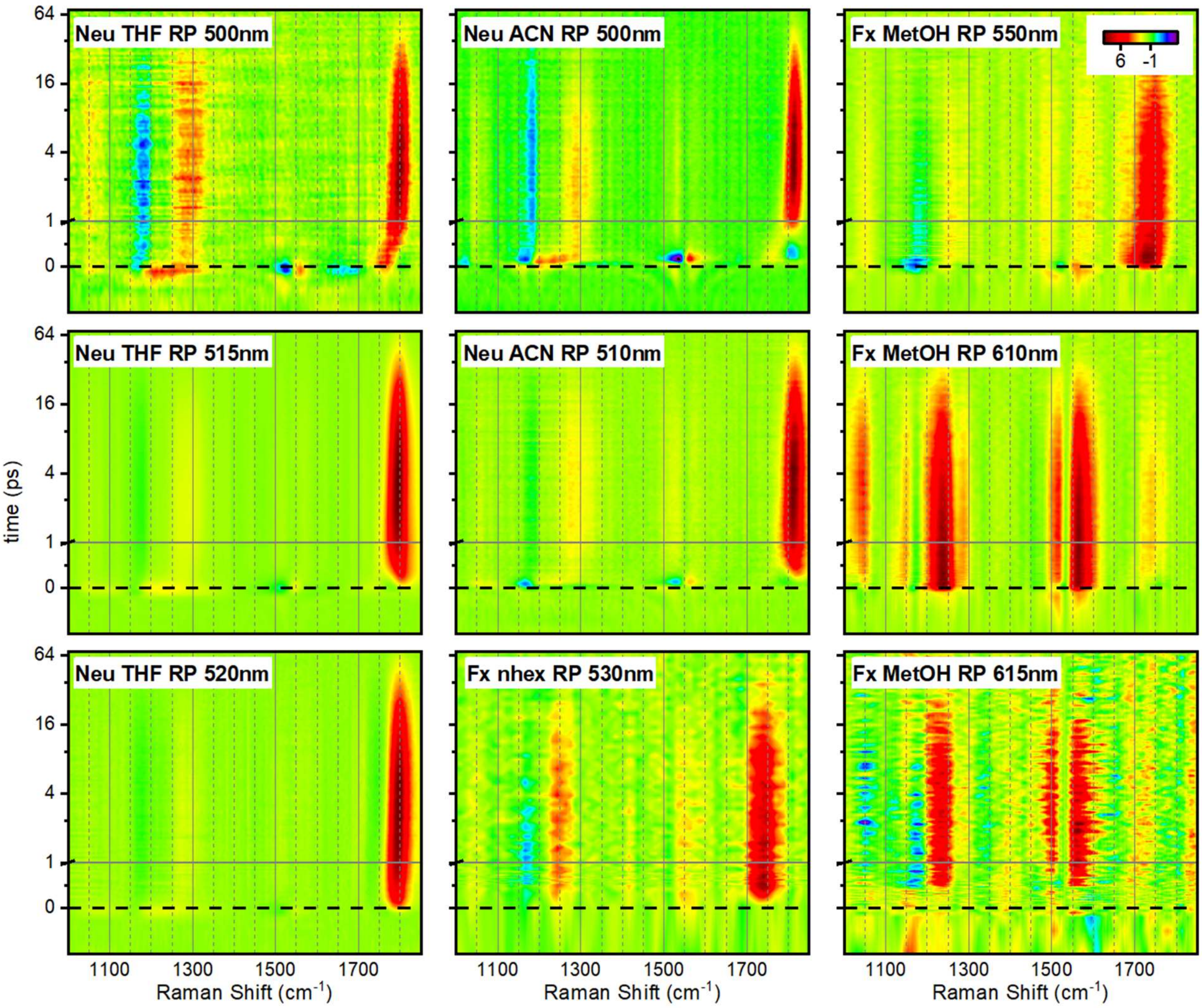

**Figure S4 | Femtosecond Stimulated Resonance Raman (FSRRS) of neurosporene in THF and Acetonitrile, and Fucoxanthin in methanol at room temperature.** Color-coded 2D time–spectral maps of the differential Raman intensity (ΔI) as a function of Raman shift ($cm^{-1}$) and pump–probe delay (linear axis to 1 ps, logarithmic thereafter), the Raman pump wavelengths are specified in each panel.

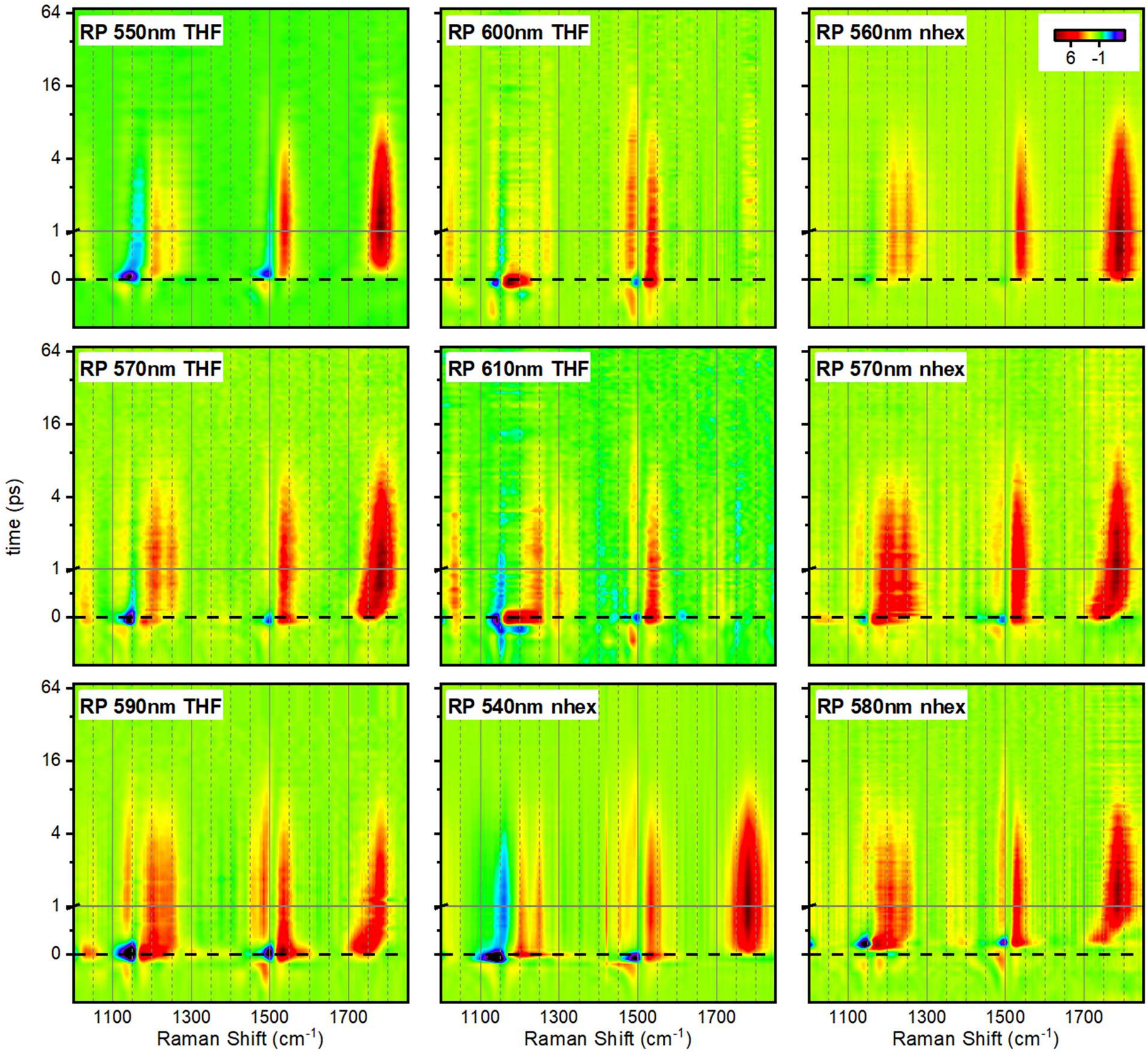

**Figure S5 | Femtosecond Stimulated Resonance Raman (FSRRS) of lycopene in THF and n-hexane at room temperature.** Color-coded 2D time–spectral maps of the differential Raman intensity (ΔI) as a function of Raman shift (cm$^{-1}$) and pump–probe delay (linear axis to 1 ps, logarithmic thereafter), recorded following actinic excitation at 510 nm; the Raman pump wavelengths are specified in each panel.

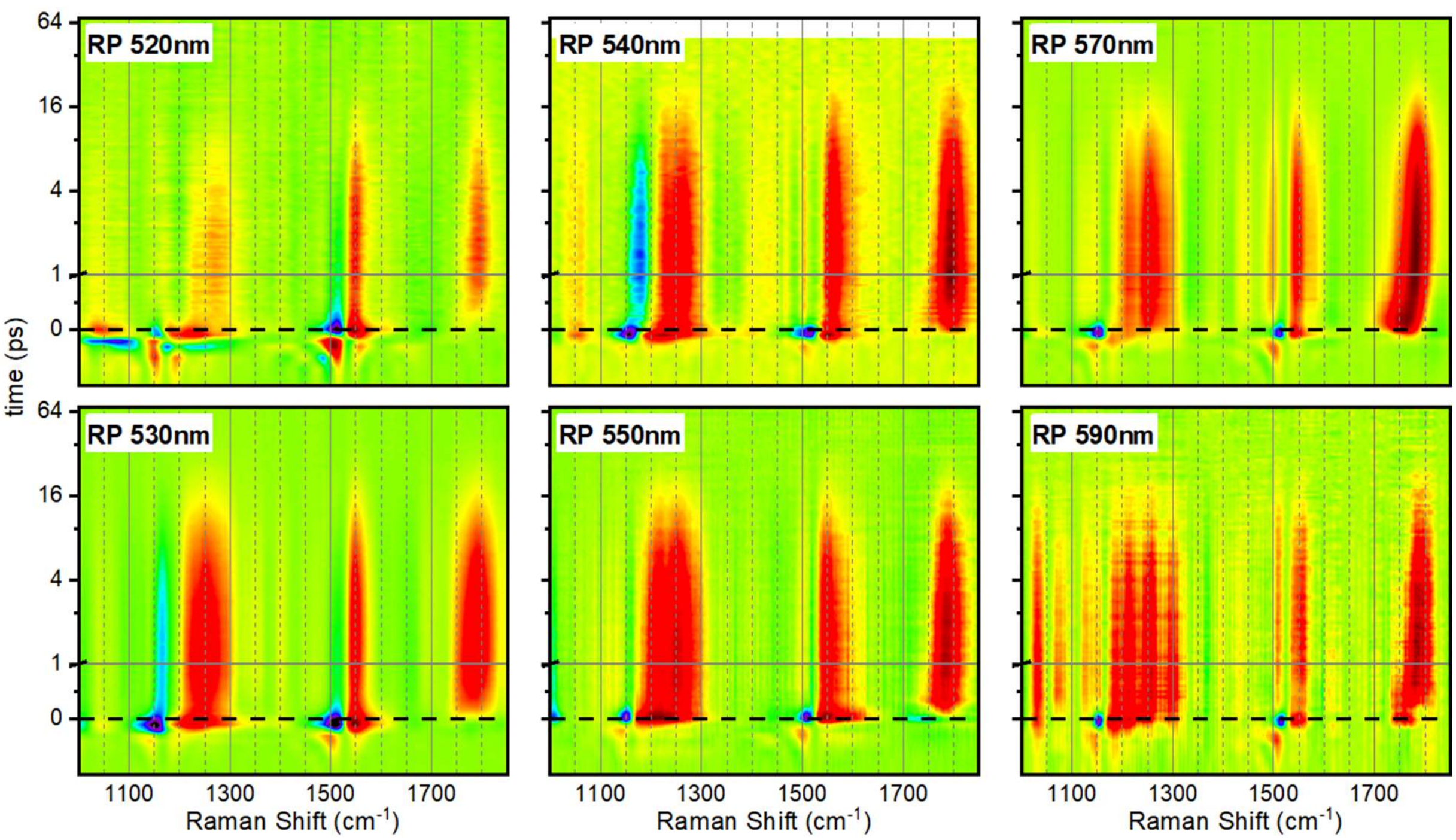

**Figure S6 | Femtosecond Stimulated Resonance Raman (FSRRS) of β-carotene in THF at 298 K.** Colour-coded time–spectral maps of the differential Raman intensity (ΔI) as a function of Raman shift ($cm^{-1}$) and pump–probe delay (linear axis to 1 ps, logarithmic thereafter), recorded following actinic pump at 510 nm; RP wavelengths are specified in each panel.

## Selection of the kinetic model

We tested three kinetic models: a 3-component sequential model, a 4-component sequential model, and a 4-component branched model without ES5. All fits were performed using global target analysis following the approach of van Stokkum et al. (*19*). The 3-component sequential model fails to reproduce the data. Both 4-component models produce mathematically acceptable fits and are shown in Figures S7–S8. However, neither model correctly disentangles the $\nu_1$c mode. As shown by the individual kinetic traces in Figure 2 of the main text, $\nu_1$c forms within the IRF (<150 fs) and decays with a lifetime distinct from both ES2 and ES3. In a sequential scheme, this mode is necessarily absorbed into ES2, where it appears as a non-evolving feature upon the ES2 → ES3 transition. Because this transition is mathematically equivalent to a sequential step, the model cannot flag the independent behaviour of $\nu_1$c. The same limitation applies to the branched model without accounting for ES5. The five-component branched model is therefore the minimal scheme that accounts for the formation and decay kinetics of all resolved $\nu_1$ components as directly observed in Figure 2.

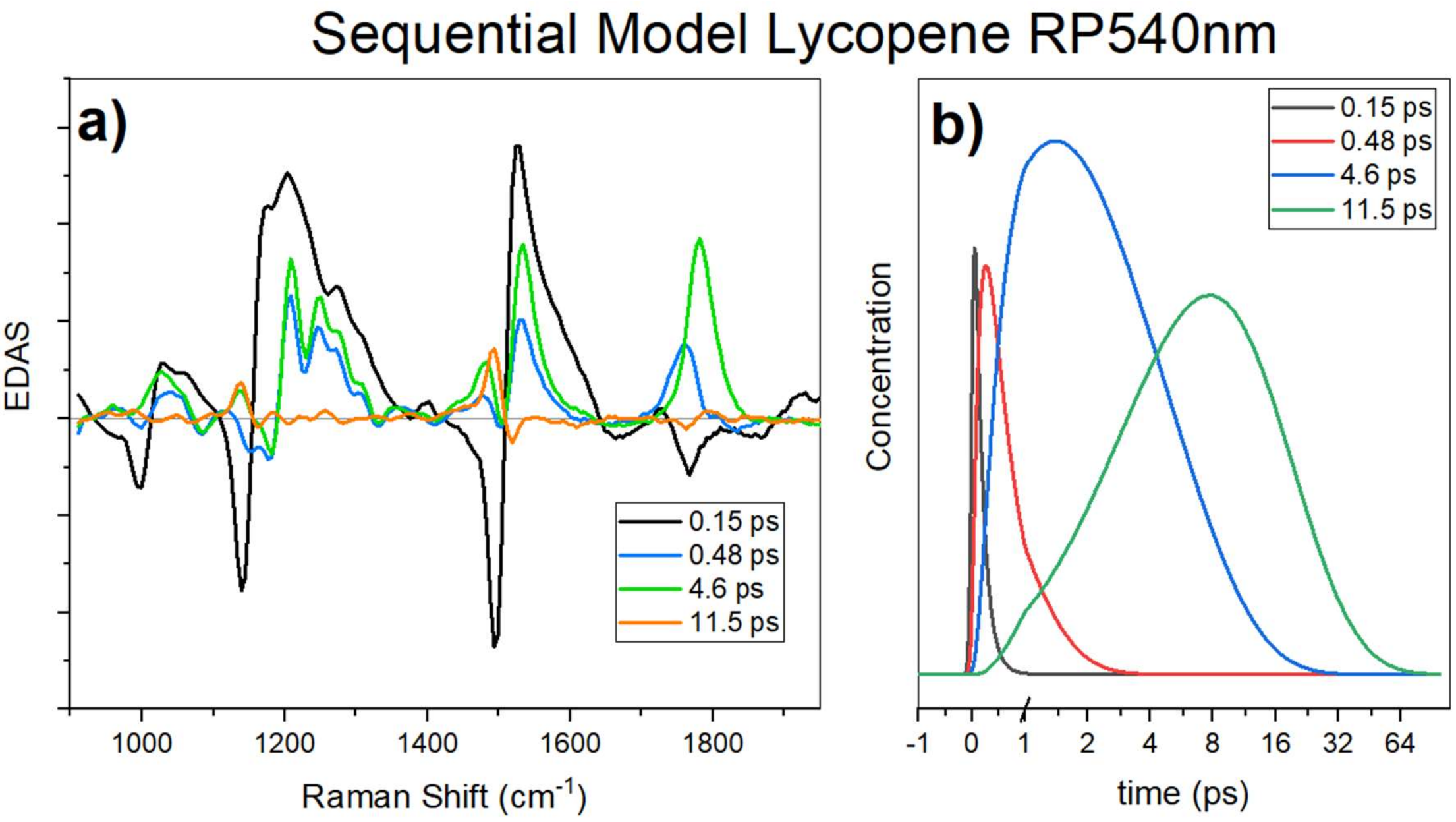

**Figure S7 |** Sequential model of 298 K FSRRS of lycopene in THF. (**a**) Evolution decay Species-associated spectra (EDAS); actinic pump 510 nm, RP 540 nm. (**b**) Concentration evolution of the species.

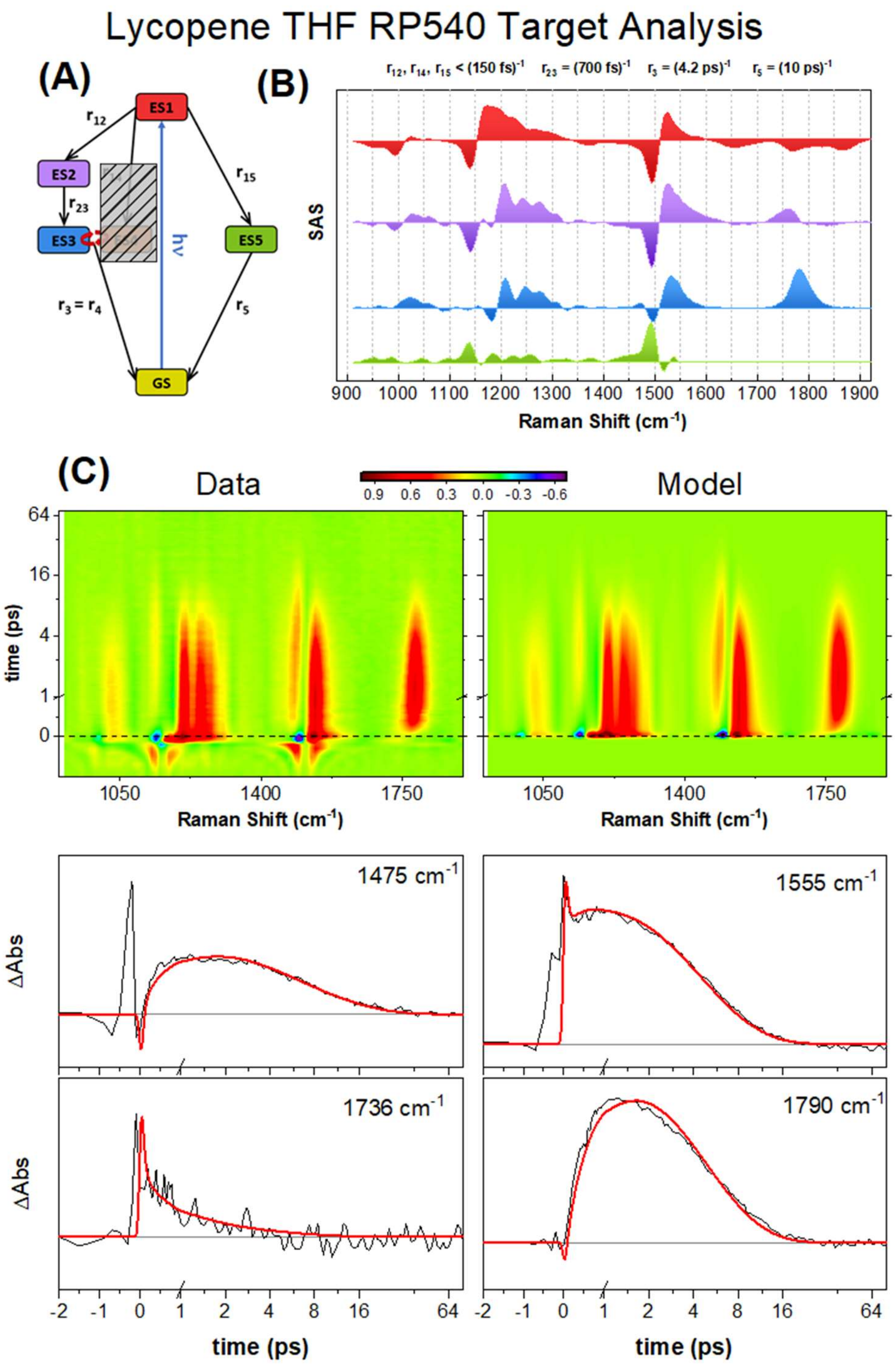

**Figure S8 | Target Analysis of 298 K FSRRS of lycopene in THF.** (A) The branched kinetic scheme with 4 components is colour-coded by state, and the fitted values are reported as lifetimes (tau = 1/r). (B) Species-associated spectra (SAS); actinic pump 510 nm, RP 540 nm. (C) Experimental traces and model obtained for the colour-coded time–spectral maps of the differential Raman intensity (ΔI) as a function of Raman shift (cm$^{-1}$) and pump–probe delay (linear axis to 1 ps, logarithmic thereafter), and kinetics and fit at each of the four C=C stretching modes.

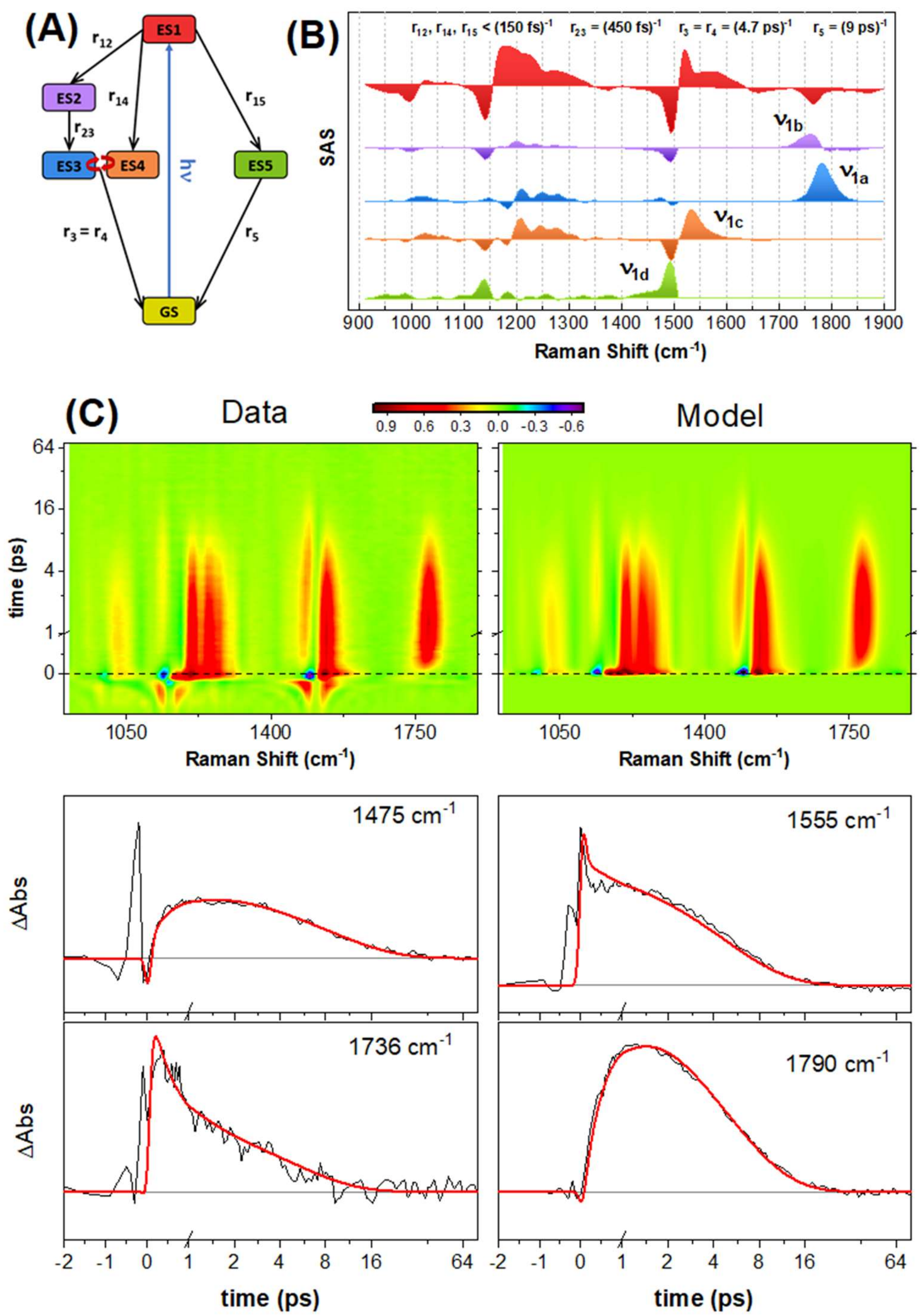

**Figure S9 | Target Analysis of 298 K FSRRS of lycopene in THF. (A)** The branched kinetic scheme with 5 components is colour-coded by state (ES1-ES5), and the fitted values are reported as lifetimes (tau = 1/r). **(B)** Species-associated spectra (SAS); actinic pump 510 nm, RP 540 nm. **(C)** Experimental traces and model obtained for the colour-coded time–spectral maps of the differential Raman intensity (ΔI) as a function of Raman shift (cm$^{-1}$) and pump–probe delay (linear axis to 1 ps, logarithmic thereafter), and kinetics and fit at each of the four C=C stretching modes.

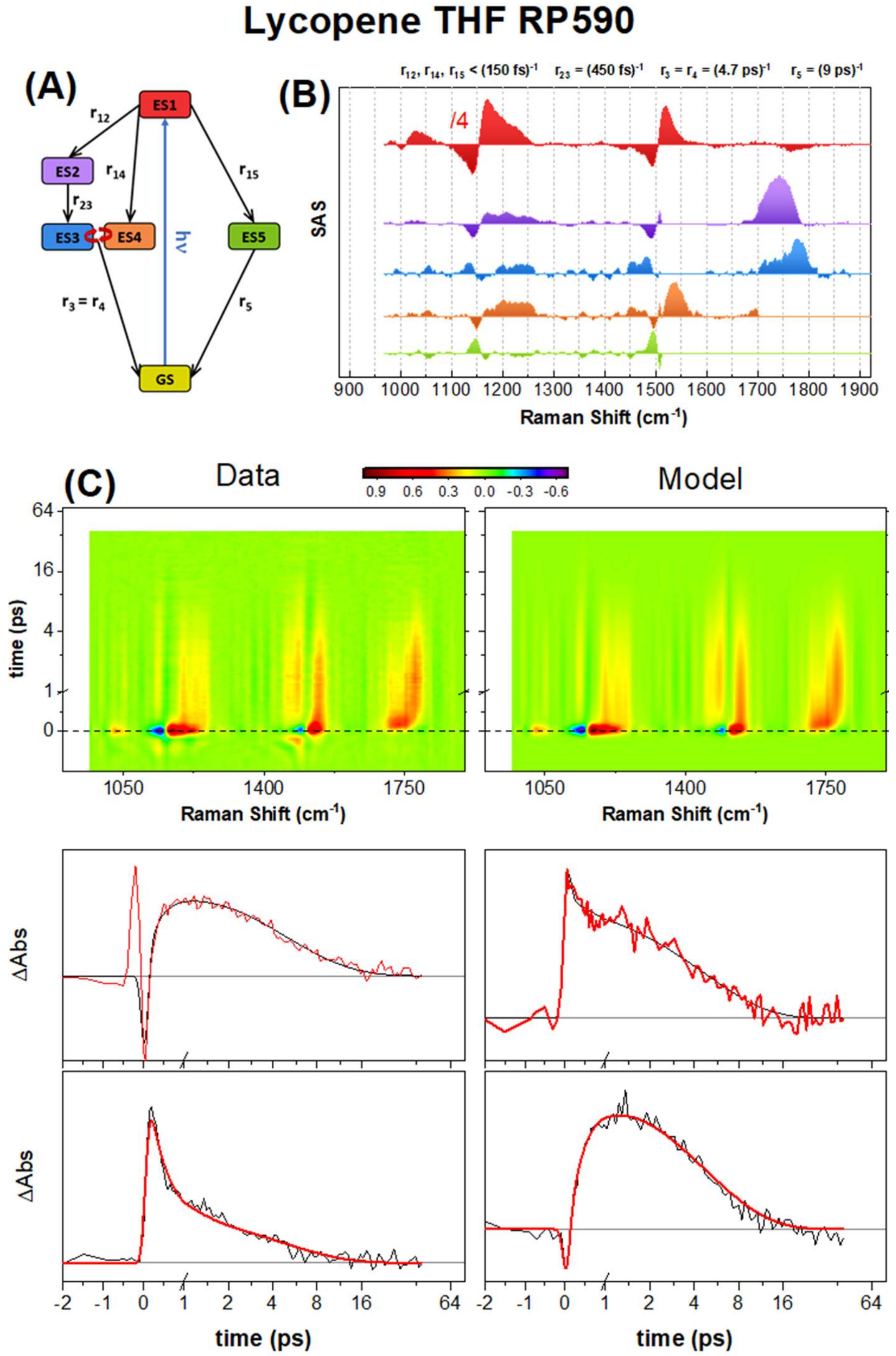

**Figure S10 | Target Analysis of 298 K FSRRS of lycopene in THF.** (A) The 5-component branched kinetic scheme is colour-coded by state (ES1-ES5), and the fitted values are reported as lifetimes (tau = 1/r). (B) Species-associated spectra (SAS); actinic pump 510 nm, RP 590 nm. (C) Experimental traces and model obtained for the colour-coded time–spectral maps of the differential Raman intensity (ΔI) as a function of Raman shift (cm$^{-1}$) and pump–probe delay (linear axis to 1 ps, logarithmic thereafter), and kinetics and fit at each of the four C=C stretching modes.

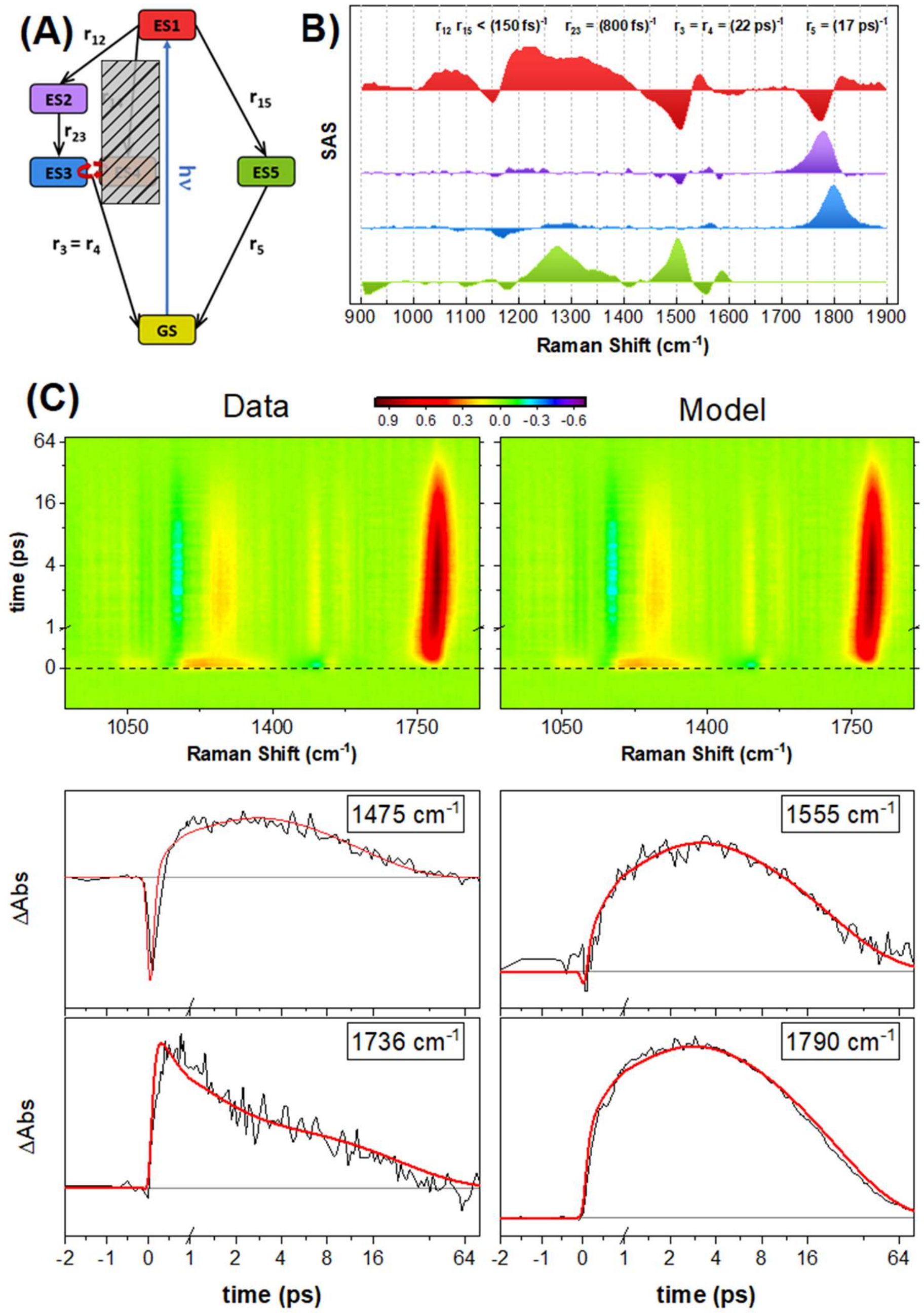

**Figure S11 | Target Analysis of 298 K FSRRS of neurosporene in THF. (A)** The branched kinetic scheme is colour-coded by state (ES1-ES5), and the fitted values are reported as lifetimes (tau = 1/r). **(B)** Species-associated spectra (SAS); actinic pump 510 nm, RP 540 nm. Note that the feature associated to ES4 ($\nu_{1c}$) is too small to be fit satisfactorily. We eliminated this parameter for this dataset, although the small vibrational mode associated to $\nu_{1c}$ is still visible in the SAS associated to ES2 and ES3. **(C)** Experimental traces and model obtained for the colour-coded time–spectral maps of the differential Raman intensity (ΔI) as a function of Raman shift (cm$^{-1}$) and pump–probe delay (linear axis to 1 ps, logarithmic thereafter), and kinetics and fit at each of the four C=C stretching modes.

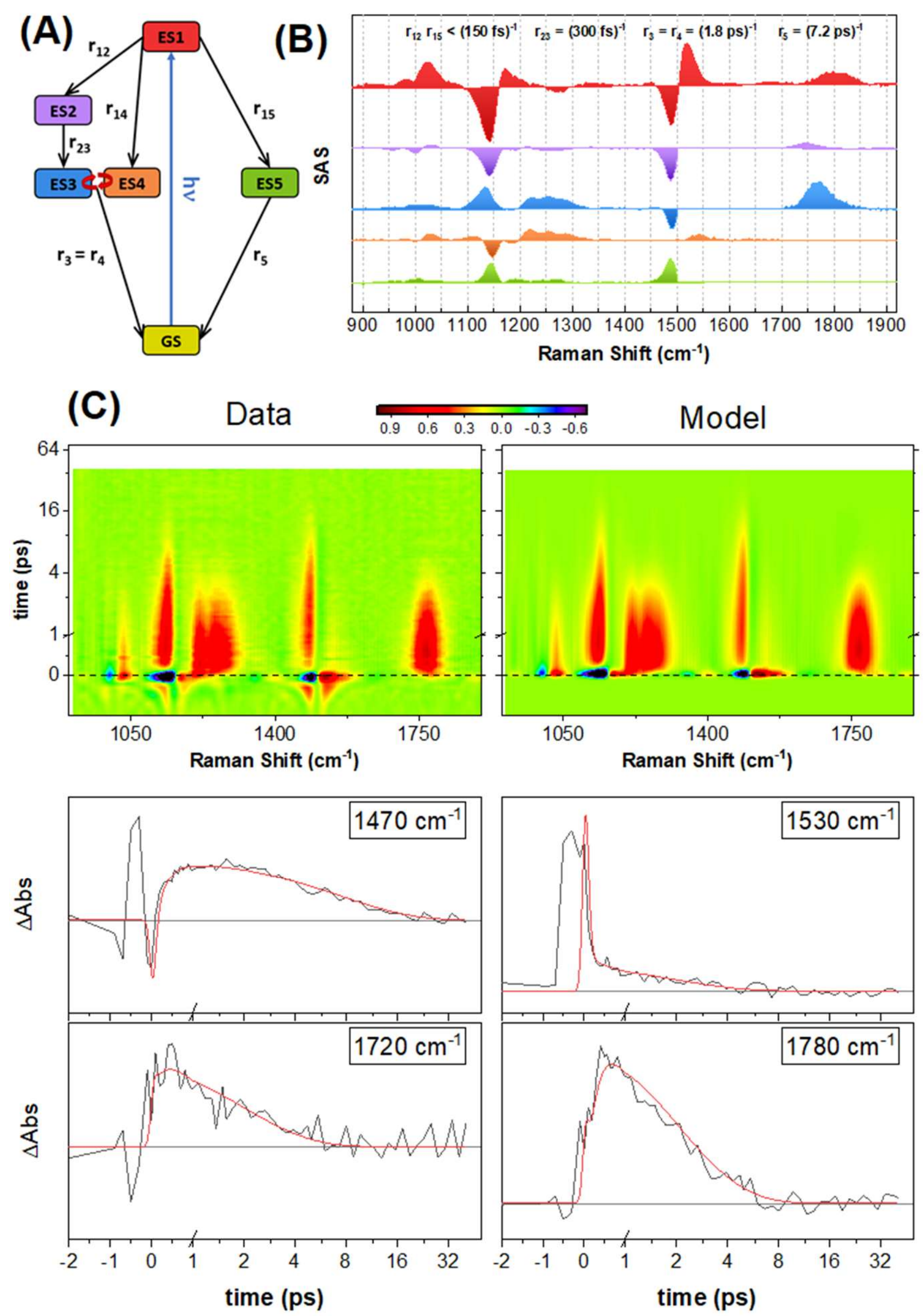

**Figure S12 | Target Analysis of 298 K FSRRS of spirilloxanthin in THF. (A)** The branched kinetic scheme is colour-coded by state (ES1-ES5), and the fitted values are reported as lifetimes (tau = 1/r). **(B)** Species-associated spectra (SAS); actinic pump 540 nm, RP 580 nm. **(C)** Experimental traces and model obtained for the colour-coded time–spectral maps of the differential Raman intensity (ΔI) as a function of Raman shift ($cm^{-1}$) and pump–probe delay (linear axis to 1 ps, logarithmic thereafter), and kinetics and fit at each of the four C=C stretching modes.

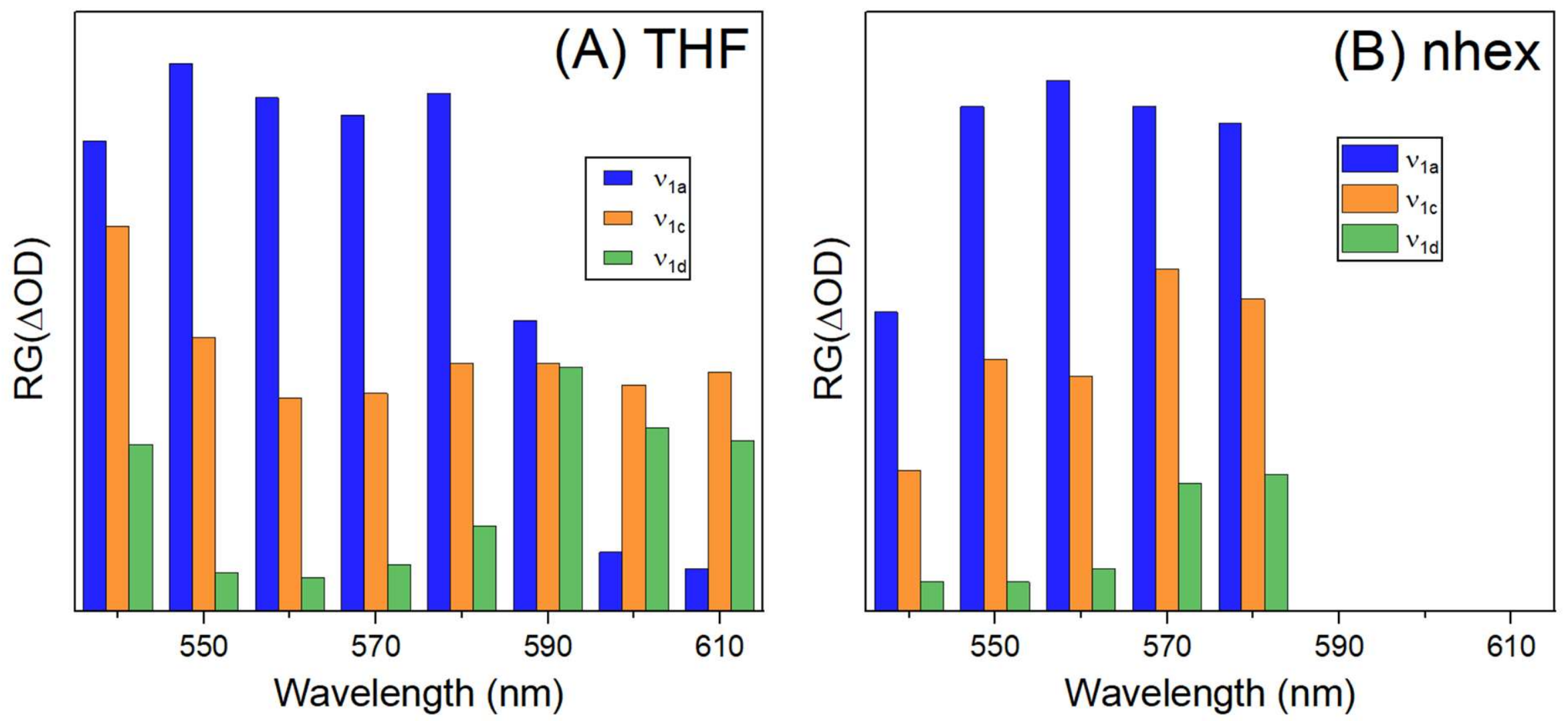

**Figure S13 |** Normalized Bar chart showing the changes of relative intensity for each Raman pump for lycopene in (A) THF, and (B) n-hexane for the C=C–stretching–mode frequencies, at $\nu_{1d}$ (green), $\nu_{1c}$ (orange), and $\nu_{1a}$ (blue).

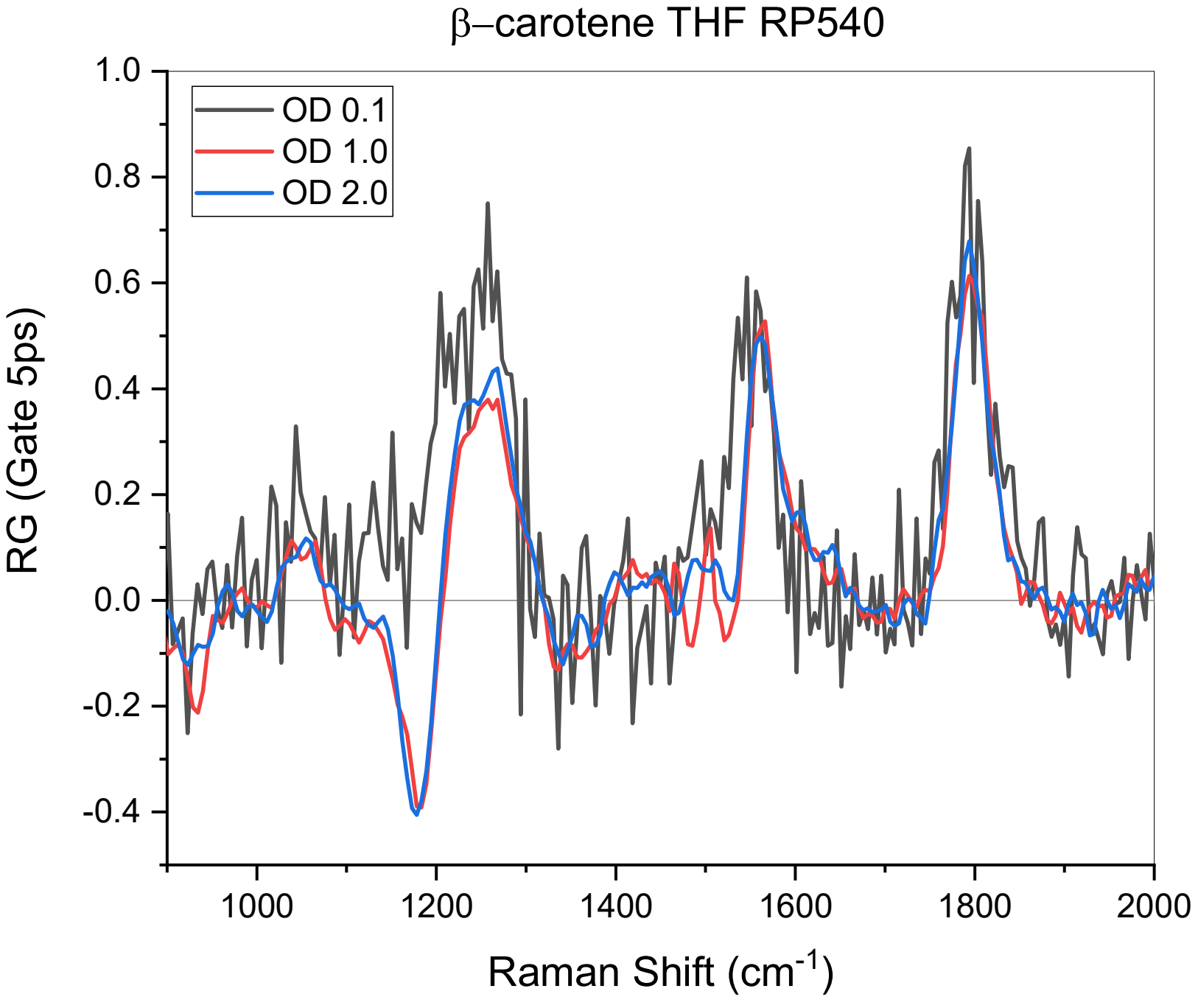

**Figure S14 |** Femtosecond stimulated resonant Raman spectra of β-carotene at room temperature with a 540 nm Raman pump. Spectra (offset for clarity) span the indicated concentrations; relative intensities of the ~1550 cm⁻¹ and ~1790 cm⁻¹ modes remain constant.

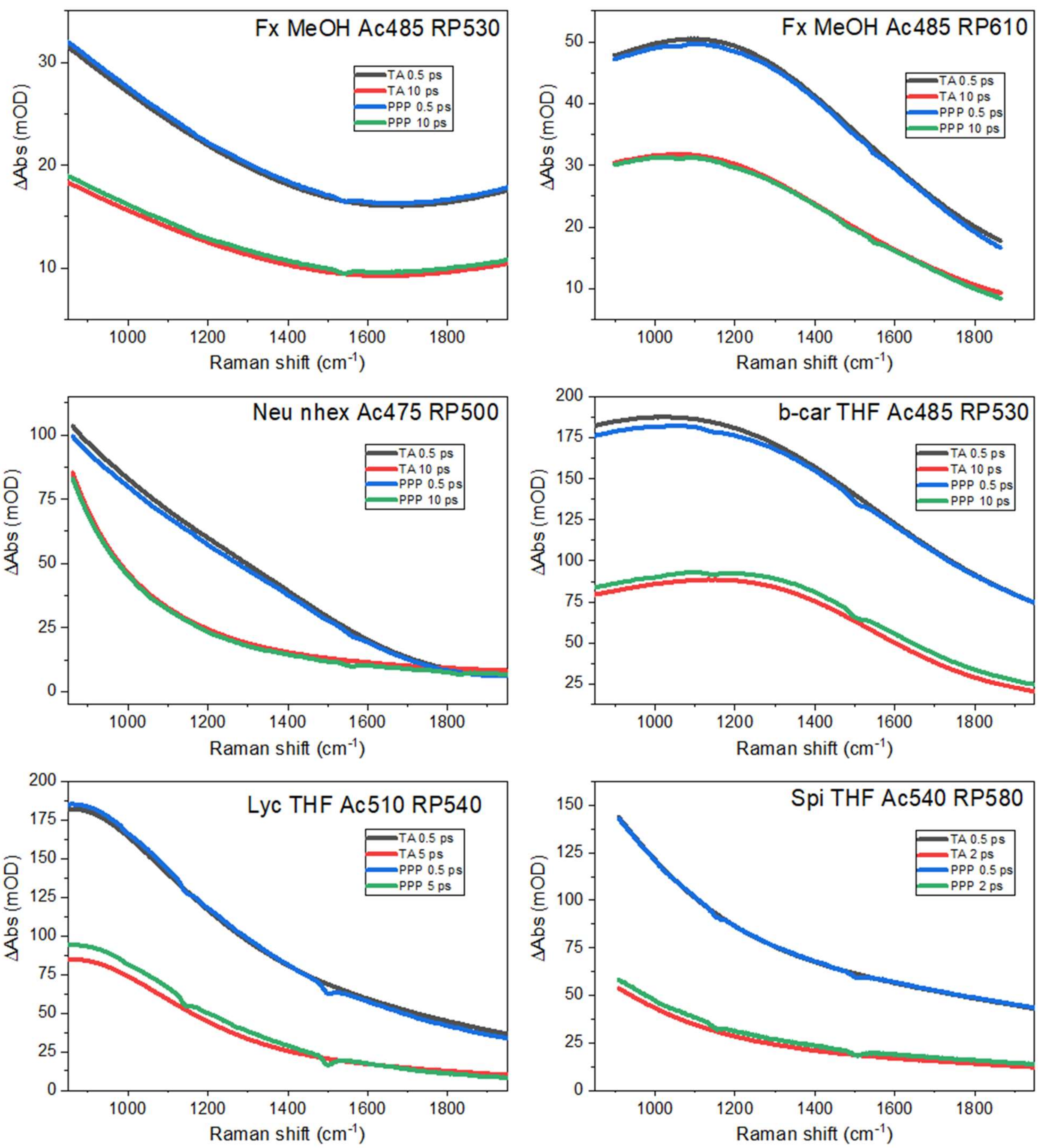

**Figure S15 |** Comparison of transient absorption and pump–pump–probe signals recorded at 0.5 ps and 5 ps for each carotenoid, with the Raman pump tuned near the resonance maximum. The Raman pump fluence is approximately 400 μJ $cm^{-2}$. The specific actinic excitation and Raman pump conditions are indicated in each panel.

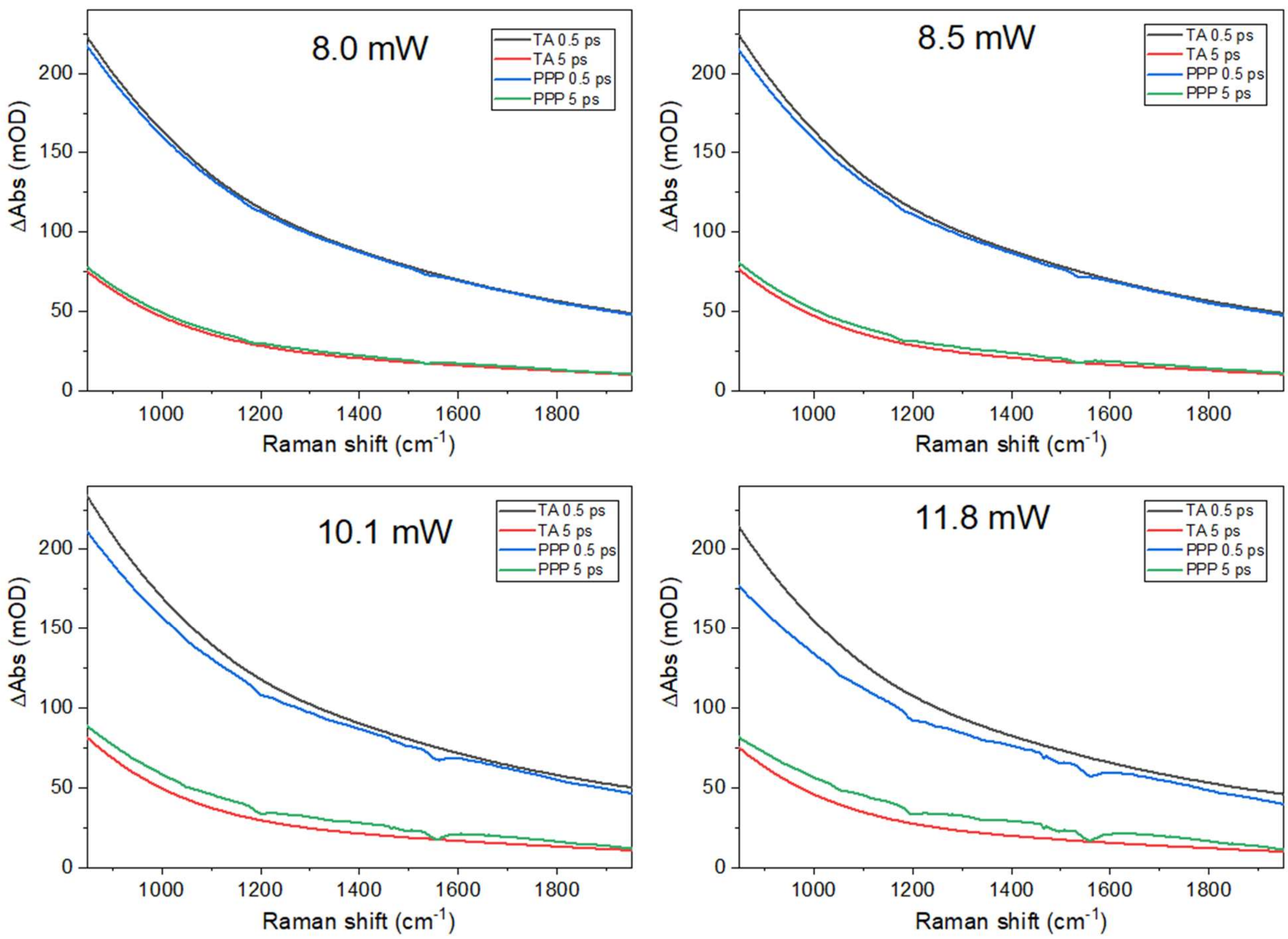

**Figure S16 |** Comparison of transient absorption and pump–pump–probe signals for lycopene in THF at room temperature recorded at 0.5 ps and 5 ps for varying Raman pump intensities, recorded following actinic excitation at 510 nm; Raman pump wavelengths 540 nm. Power dependence from 8.0 to 11.8 mW, equivalent to 400 - 600 µJ cm$^{-2}$ fluence.

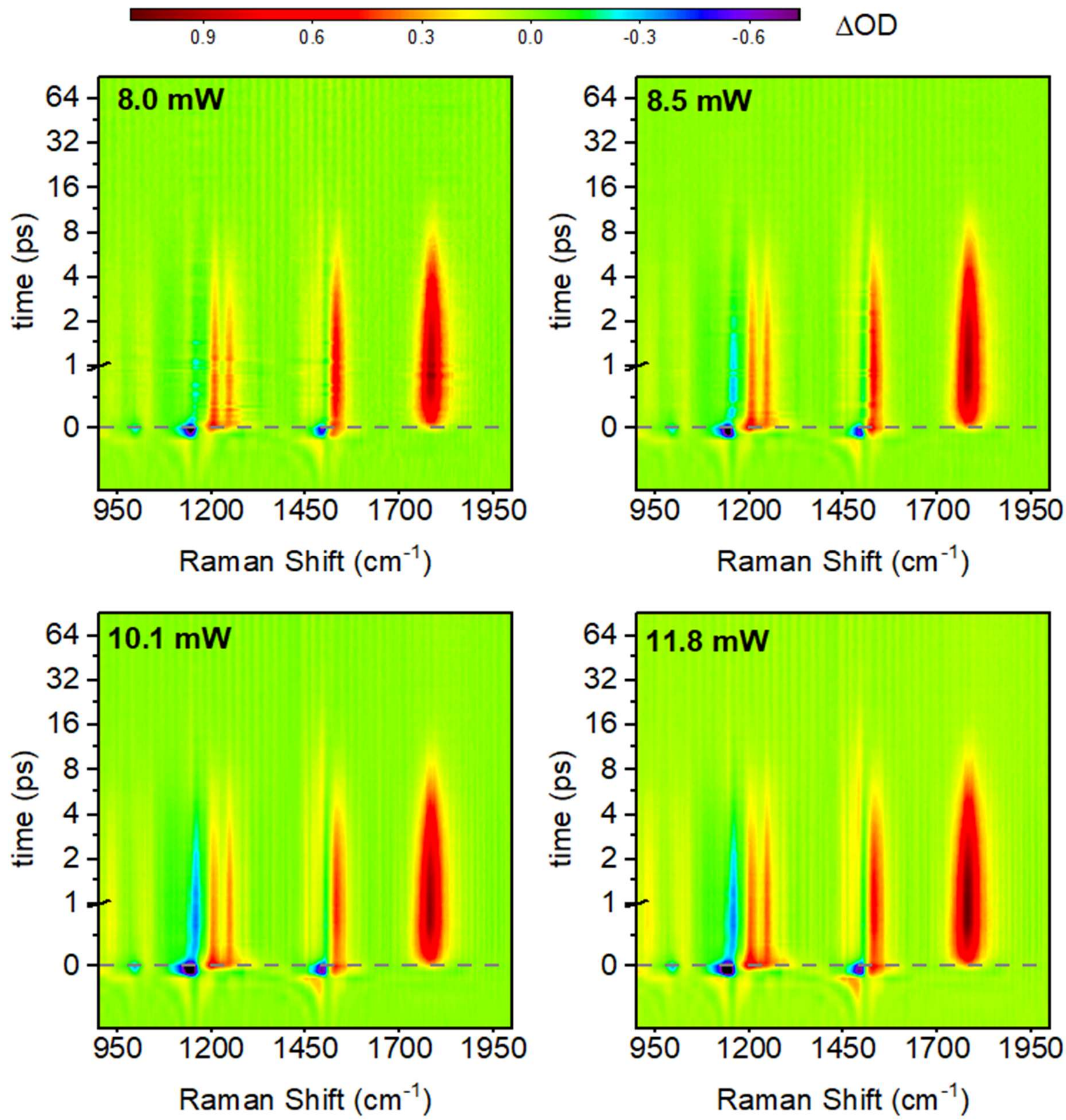

**Figure S17 |** Power dependence of Femtosecond Stimulated Resonant Raman of lycopene in THF at room temperature, from 8.0 to 11.8 mW (equivalent to 400 - 600 μJ cm$^{-2}$ fluence). Color-coded 2D time–spectral maps of the differential Raman intensity (ΔI) as a function of Raman shift (cm$^{-1}$) and pump–probe delay (linear axis to 1 ps, logarithmic thereafter), recorded following actinic excitation at 510 nm; Raman pump wavelength 540 nm.

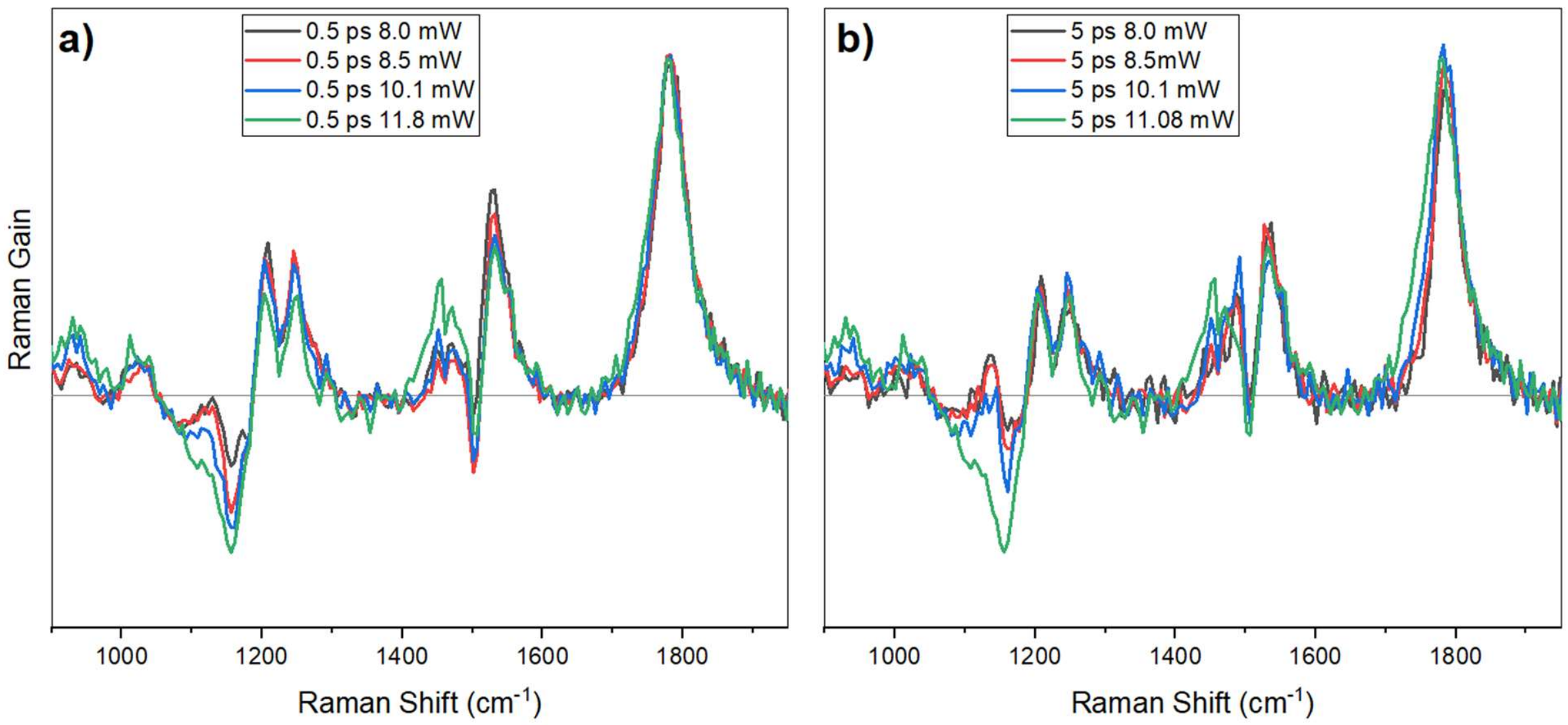

**Figure S18 |** Power dependence of Femtosecond Stimulated Resonant Raman of lycopene in THF at room temperature, from 8.0 to 11.8 mW (equivalent to 400 - 600 μJ cm$^{-2}$ fluence). Gated spectra at 0.5 ps and 5 ps of the differential Raman intensity (ΔI) as a function of Raman shift (cm$^{-1}$) and pump–probe delay (linear axis to 1 ps, logarithmic thereafter), recorded following actinic excitation at 510 nm; Raman pump wavelengths 540 nm.

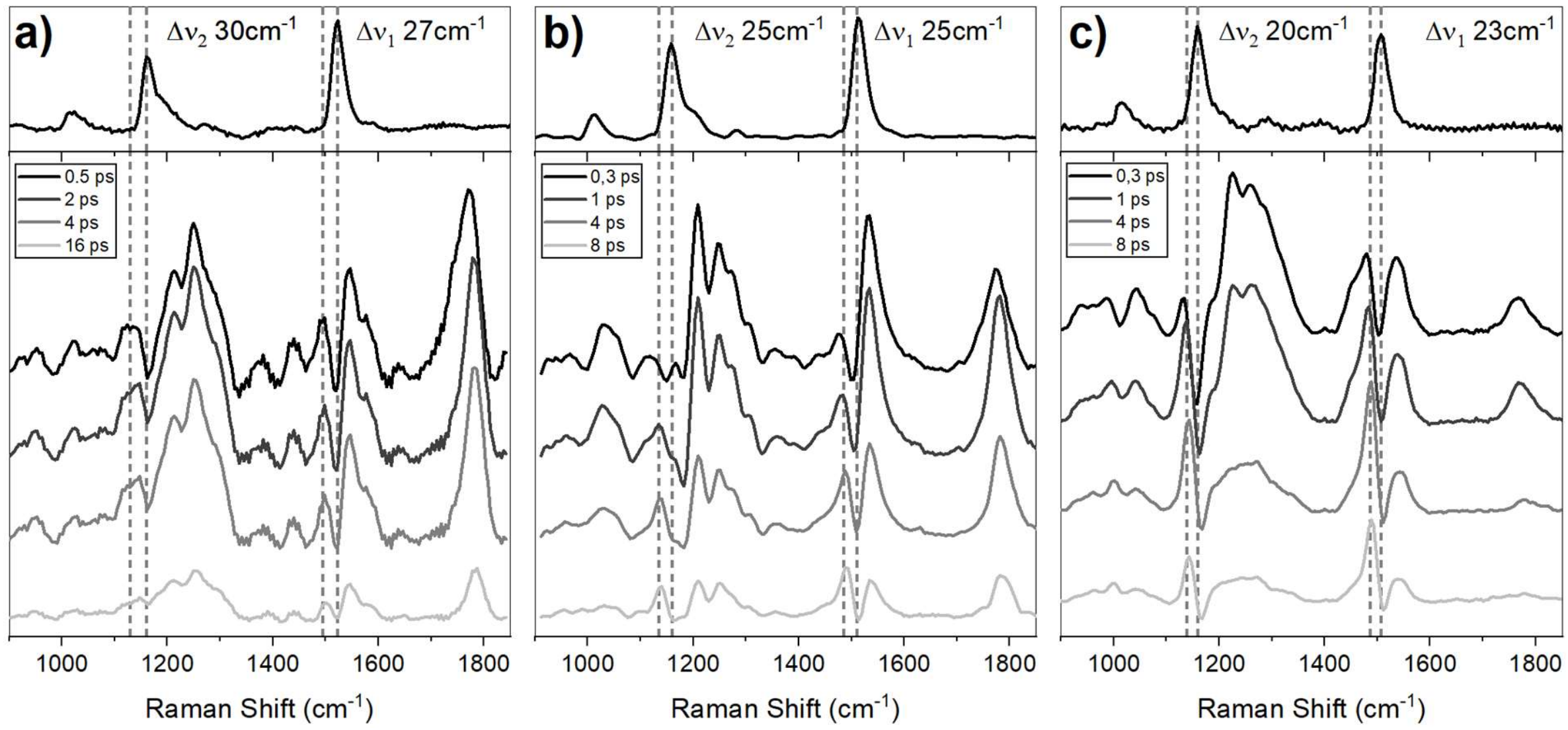

**Figure S19 |** Time-gated Raman spectra of (a) β-carotene (RP 570 nm), (b) lycopene (RP 540 nm), and (c) spirilloxanthin (RP 560 nm), together with ground-state Raman spectra. Dashed grey lines indicate the frequency shifts between ground-state modes and those associated with the S* state.